\documentclass{article}
\usepackage[a4paper, total={6in, 9in}]{geometry}
\usepackage[utf8]{inputenc}
\usepackage{amsmath,amsfonts}
\usepackage{booktabs} 
\usepackage{caption} 
\usepackage{subcaption} 
\usepackage{graphicx}
\usepackage{pgfplots}
\usepackage[all]{nowidow}
\usepackage[utf8]{inputenc}
\usepackage{tikz}
\usetikzlibrary{er,positioning,bayesnet}
\usepackage{multicol}
\usepackage{algpseudocode,algorithm,algorithmicx}

\definecolor{blue}{HTML}{1F77B4}
\definecolor{orange}{HTML}{FF7F0E}
\definecolor{green}{HTML}{2CA02C}

\pgfplotsset{compat=1.14}

\begin{document}
\title{ High-Frequency Instabilities of Stokes Waves}
%
%
\author{\normalsize Ryan P. Creedon$^1$, Bernard Deconinck$^2$, Olga Trichtchenko$^3$}
%
%

\date{\normalsize July 23, 2021}

\maketitle              
\vspace*{-0.7cm}
{\begin{center} {\scriptsize\noindent $^1$Department of Applied Mathematics, University of Washington, Seattle, WA, USA, {creedon@uw.edu}\\ \vspace*{-0.0cm} \noindent $^2$Department of Applied Mathematics, University of Washington, Seattle, WA, USA, {deconinc@uw.edu} \\ \vspace{-0.125cm}\noindent $^3$Department of Physics and Astronomy, The University of Western Ontario, London, ON, CA, {otrichtc@uwo.ca}}\\~\\~\
\emph{This paper is dedicated to Harvey Segur, on the occasion of his 80th birthday.} \\~\vspace*{-0.325cm}\end{center}}

\begin{abstract}
Euler's equations govern the behavior of gravity waves on the surface of an incompressible, inviscid, and irrotational fluid of arbitrary depth. We investigate the spectral stability of sufficiently small-amplitude, one-dimensional Stokes waves, {\em i.e.}, periodic gravity waves of permanent form and constant velocity, in both finite and infinite depth.
We develop a perturbation method to describe the first few high-frequency instabilities away from the origin, present in the spectrum of the linearization about the small-amplitude Stokes waves. 
Asymptotic and numerical computations of these instabilities are compared for the first time to excellent agreement. 
\end{abstract}
\section{Introduction}

\begin{figure}[tb]
\centering
    \includegraphics[height=3.2cm,width=12.2cm]{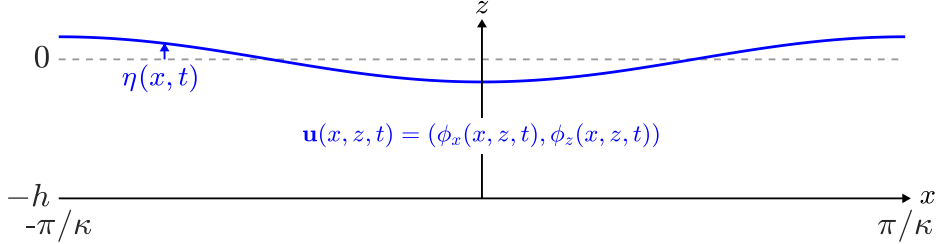}
    \caption{A schematic of 1D gravity waves in finite depth $h$.  In this work, the surface displacement $\eta$ and velocity field  ${\bf u} = \left(\phi_x,\phi_z\right)$ are $2\pi/\kappa$-periodic in the $x$-direction.\\}   \label{fig1}
\end{figure}

We consider periodic gravity waves along a 1D surface of an incompressible, inviscid, and irrotational fluid of arbitrary depth. These waves are governed by Euler's equations \cite{euler1757a}-\cite{euler1761}
\begin{subequations}
\begin{align}
\phi_{xx} + \phi_{zz} &= 0 \label{1a} \quad \hspace*{.2cm} \textrm{in} \quad \{(x,z):|x|<\pi/\kappa ~\textrm{and} -h < z < \eta \},  \\
    \eta_t + \eta_x \phi_x &= \phi_z \quad \textrm{on} \quad z=\eta, \\ 
    \phi_t + \tfrac{1}{2} \left(\phi_x^2 + \phi_z^2\right) + g\eta &= 0 \quad \hspace*{.2cm} \textrm{on} \quad z=\eta, \\
    \phi_z &= 0, \quad \hspace*{0.07cm} \textrm{on} \quad z=-h, \label{1d}
    \end{align}
\end{subequations}
and satisfy the periodicity conditions
\begin{subequations}
\begin{align}
    \eta(-\pi/\kappa,t) &= \eta(\pi/\kappa,t), \label{2a} \\ \quad \phi_x(-\pi/\kappa,z,t) = \phi_x(\pi/\kappa,z,t), &\quad \phi_z(-\pi/\kappa,z,t) = \phi_z(\pi/\kappa,z,t). \label{2b}
\end{align}
\end{subequations}
In these equations, $\eta=\eta(x,t)$ is the surface displacement of the fluid, $\phi=\phi(x,z,t)$ is the velocity potential inside the bulk of the fluid, $g$ is the acceleration due to gravity, $h$ is the depth of the fluid, and $\kappa$ is the wavenumber of the surface displacement, see Figure \ref{fig1}. Subscripts $x$ and $t$ denote partial differentiation. 

Stokes \cite{stokes1847} showed in 1847 that periodic, traveling-wave solutions of \eqref{1a}-\eqref{1d} in infinite depth can be expressed as a power series in a small-parameter $\varepsilon$ that scales with the amplitude of the waves. Nekrasov \cite{nekrasov21} proved the convergence of this series in 1921, and the works of Levi-Civita \cite{levicivita25} and Struik \cite{struik26} extended these considerations to the case of finite depth, see Section 3 and Appendix A for more details.

The stability of Stokes waves with respect to longitudinal perturbations was first studied in the 1960s by Benjamin \emph{\&} Feir \cite{benjamin67, benjaminfeir67} and Whitham \cite{whitham67}. These independent investigations concluded that Stokes waves are modulationally unstable, provided $\kappa h > 1.3627...$. This is now referred to as the Benjamin-Feir instability. The presence of this instability was proven rigorously in finite depth by Bridges \emph{\&} Mielke \cite{bridgesmielke95} and in infinite depth by Nguyen \emph{\&} Strauss \cite{nguyenstrauss20}.

In the 1970s, Bryant \cite{bryant74}-\cite{bryant78} studied the stability of Stokes waves with respect to co-periodic and transverse perturbations in shallow depth ($\kappa h < 1.3627...$), while Longuet-Higgins \cite{longuethiggins78a, longuethiggins78b} considered infinite depth with longitudinal perturbations that were sub- and super-harmonic to the fundamental period of the Stokes wave. 
McLean \cite{mclean82} extended this work to finite depth and transverse perturbations. Over the next decades, several papers focused on the transverse instability of Stokes waves \cite{akersnicholls12,franciuskharif06,kharif90}, see also \cite{craik04,grimshaw05,yuenlake80}.

 In 2009, using a reformulation of Euler's equations developed by Ablowitz \emph{et al.} \cite{ablowitzetal06}, Deconinck \emph{\&} Oliveras \cite{deconinckoliveras11} numerically revisited the stability of Stokes waves with respect to quasi-periodic perturbations (parameterized by a Floquet exponent $\mu \in \mathbb{R}$), encompassing both super- and sub-harmonic perturbations. 
This results in a spectral problem that has a countable number of finite-multiplicity eigenvalues for each value of the Floquet exponent \cite{kapitulapromislow13}. These eigenvalues control the exponential growth rates of the perturbations, and the union of these point spectra defines the  stability spectrum of the Stokes waves, to be more precisely defined in Section 4 of this paper.

The stability spectrum depends analytically on the amplitude $\varepsilon$ of the Stokes waves \cite{nicholls07}. In addition, for fixed $\varepsilon$, the spectrum is symmetric with respect to the real and imaginary axes, since \eqref{1a}-\eqref{1d} is Hamiltonian \cite{zakharov68}. Thus, Stokes waves are spectrally stable only when the stability spectrum is a subset of the imaginary axis. Otherwise, there exists a Floquet exponent and corresponding eigenvalue for which the perturbation grows in time.
 
\begin{figure}[tb]
    \centering
    \includegraphics[height=12.5cm,width=11.5cm]{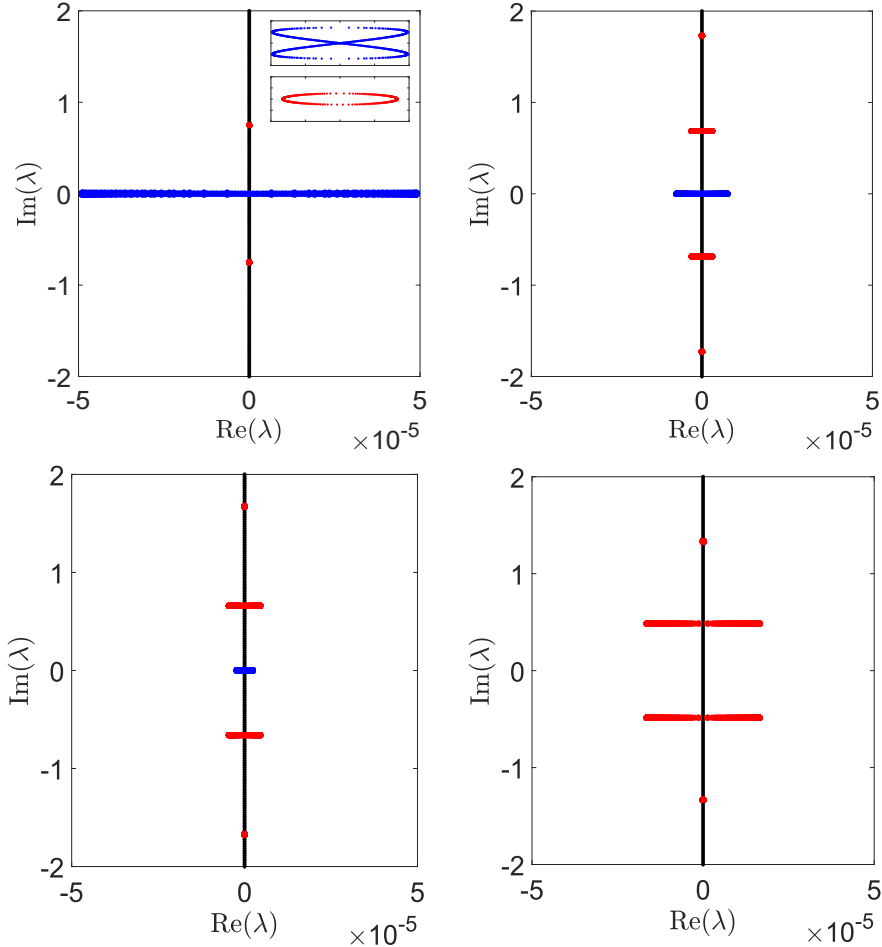}
    \caption{The stability spectrum of a $2\pi$-periodic Stokes wave with amplitude $\varepsilon = 0.01$ and (Top, Left) $h = \infty$, (Top, Right) $h=1.5$, (Bottom, Left) $h=1.4$, and (Bottom, Right) $h=1$. The Benjamin-Feir figure-eight is colored blue. The high-frequency isolas are colored red. Purely imaginary eigenvalues are colored black. A zoom-in of the Benjamin-Feir and high-frequency instabilities are inlaid in the top, left plot.\\}
    \label{fig2}
\end{figure}

In Figure \ref{fig2}, we use the Floquet-Fourier-Hill (FFH) method \cite{curtisdeconinck10,deconinckkutz06} to compute stability spectra of $2\pi$-periodic Stokes waves with amplitude $\varepsilon = 0.01$ in various depths. When $\kappa h > 1.3627...$, we observe the Benjamin-Feir instability as a figure-eight pattern at the origin. We also find unstable eigenvalues away from the origin, referred to as high-frequency instabilities. Unlike the Benjamin-Feir instability, high-frequency instabilities appear in the stability spectrum for all values of $\kappa h$. They even dominate the Benjamin-Feir instability when $1.3627...<\kappa h < 1.4305...$ \cite{deconinckoliveras11}. The topic of this paper is the study of these high-frequency instabilities using formal perturbation methods, as described below. 

High-frequency instabilities develop from a Hamiltonian-Hopf bifurcation: a nonzero, repeated eigenvalue $\lambda_0$ of the zero-amplitude stability spectrum ($\varepsilon=0$) \cite{akersnicholls12,deconincktrichtchenko17,mackaysaffman86} leaves the imaginary axis as the amplitude increases. When $0<\varepsilon \ll 1$, a connected locus of unstable eigenvalues forms, which we call a high-frequency isola (red inset in Figure \ref{fig2}). The isola is parameterized by values of $\mu$ near $\mu_0$, the Floquet exponent corresponding to $\lambda_0$.

\begin{figure}[tb]
    \centering
    \includegraphics[height=6.5cm,width=14cm]{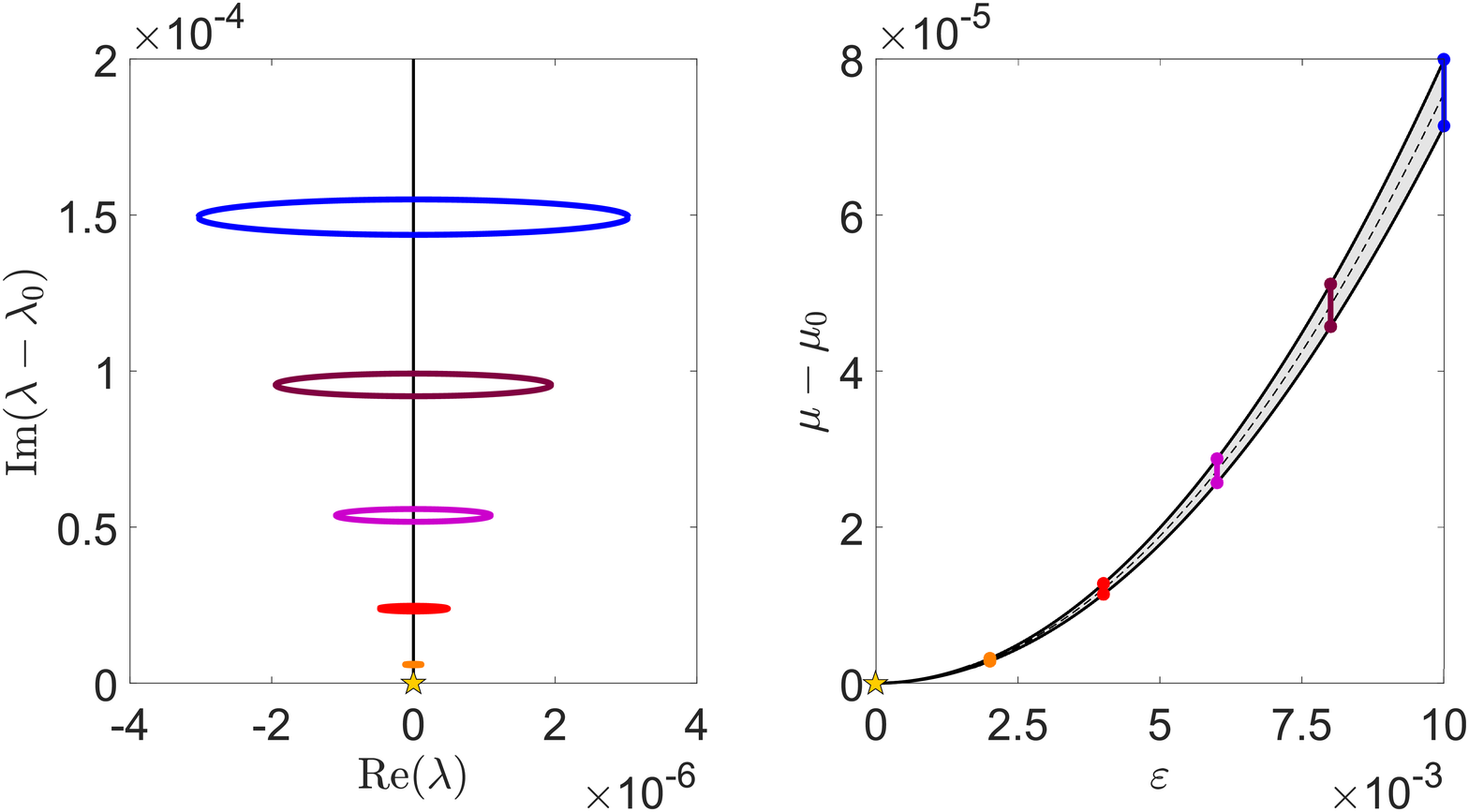}
    \caption{(Left) The high-frequency isola closest to the origin for a $2\pi$-periodic Stokes wave in depth $h = 1.5$ with amplitude $\varepsilon = 2 \times 10^{-3}$ (orange), $\varepsilon = 4 \times 10^{-3}$ (red), $\varepsilon = 6 \times 10^{-3}$ (magenta), $\varepsilon = 8 \times 10^{-3}$ (purple), and $\varepsilon = 10^{-2}$ (blue). The imaginary axis is recentered to show the drift of the isola from the collided eigenvalues at $\lambda_0$. The isolas are computed using the perturbation method developed in this paper. (Right) The interval of Floquet exponents that parameterizes the isola closest to the origin in depth $h=1.5$ as a function of the amplitude. The solid black lines indicate the boundaries of this interval, while the dashed black line gives the Floquet exponent corresponding to the most unstable eigenvalue on the isola. The colored lines give the Floquet exponents corresponding to the similarly colored isolas in the left figure. The Floquet axis is recentered to show the drift of the parameterizing interval from the Floquet exponent $\mu_0$ that corresponds to the collided eigenvalues. The paramaterizing interval is also computed using the perturbation method in this paper.\\}
    \label{fig3}
\end{figure}

High-frequency isolas are challenging to detect for numerical methods like FFH as they exist for narrow, specific ranges of the Floquet exponent. 
To complicate matters further, this narrow interval of Floquet exponents drifts from $\mu_0$ as $\varepsilon$ increases. In most depths, $\mu_0$ is no longer within the interval that parameterizes the first high-frequency isola for small, positive values of $\varepsilon$. Therefore, to capture an isola using numerical methods, one must not only take into account the narrow interval of Floquet exponents that parameterizes the isola, but also its drift from $\mu_0$ as $\varepsilon$ changes (Figure \ref{fig3}).

In this paper, we derive formal asymptotic expressions for isolas close to the origin, both in finite and infinite depth. Specifically, for each isola we derive
\begin{enumerate}
    \item[(i)] an interval of Floquet exponents that is asymptotic to the interval parameterizing the isola,
    \item[(ii)] an asymptotic expansion for the most unstable eigenvalue on the isola, and 
    \item[(iii)] a closed-form expression for the curve asymptotic to the isola.
\end{enumerate} 

Our asymptotic expressions are compared directly with numerical results of the FFH method. For almost all $\kappa h$ (except a few isolated values), our asymptotic expressions predict that Stokes waves of sufficiently small (but finite) amplitude are unstable with respect to high-frequency instabilities, extending recent work by Hur \emph{\&} Yang \cite{huryang20} that establishes the instability closest to the origin only for $\kappa h\in (0.86430..., 1.00804...)$, see Section~5. 

Our approach is an extension of standard eigenvalue perturbation theory \cite{kato66}, as we crucially let the Floquet exponent depend on the wave amplitude $\varepsilon$ to account for the drift in the isola's Floquet parameterization. This same approach was first used in Creedon \emph{et al.} \cite{creedonetal21a} on the Kawahara equation and in Creedon \emph{et al.} \cite{creedonetal21b} on a Boussinesq-Whitham system. An outline of the leading-order calculations of the method in infinite depth is also used by Akers \cite{akers15}, where the emphasis is on understanding the analyticity properties of the stability spectrum as a function of the boundary conditions imposed on the perturbations ({\em i.e.}, as a function of the Floquet exponent), and on the connections with resonant interaction theory. 


\section{The AFM Formulation}

Euler's equations \eqref{1a}-\eqref{1d} together with the auxiliary conditions \eqref{2a}-\eqref{2b} constitute a boundary value problem for Laplace's equation in a domain evolving nonlinearly in time. Depending on the application, other formulations of gravity waves may be preferred over \eqref{1a}-\eqref{1d}. 
We consider the Ablowitz-Fokas-Musslimani (AFM) formulation, first proposed in \cite{ablowitzetal06}. This formulation has dependence only on surface variables, as in Zakharov \cite{zakharov68} or Craig \emph{\&} Sulem \cite{craigsulem93}, but avoids direct numerical computations of the Dirichlet-to-Neumann operator. 

As shown in \cite{ablowitzhaut08,oliveras09}, Euler's equations \eqref{1a}-\eqref{1d} with the lateral periodic boundary conditions \eqref{2a}-\eqref{2b} are equivalent to the following system for the surface variables $\eta$ and $q = \phi(x,\eta,t)$:
\begin{subequations}
\begin{align}
    \int_{-\pi/\kappa}^{\pi/\kappa} e^{-i\kappa m x}\Big[\eta_t\cosh\left(\kappa m\left(\eta + h\right) \right) +iq_x\sinh\left(\kappa m\left(\eta + h \right) \right) \Big] dx &= 0, \quad m \in \mathbb{Z} \setminus \{0\}, \label{4a} \\
    q_t + \frac12 q_x^2 +g\eta - \frac{1}{2}\frac{\left(\eta_t +\eta_xq_x\right)^2}{1+\eta_x^2} &= 0. \label{4b}
\end{align}
\end{subequations}
We call \eqref{4a} and \eqref{4b} the nonlocal and local equations of the AFM formulation, respectively. 

We write \eqref{4a}-\eqref{4b} in a traveling frame $x \rightarrow x -ct$:
\begin{subequations}
\begin{align}
    \int_{-\pi/\kappa}^{\pi/\kappa} e^{-i\kappa m x}\Big[\left(\eta_t-c\eta_x\right)\cosh\left(\kappa m\left(\eta + h\right) \right) +iq_x\sinh\left(\kappa m\left(\eta + h \right) \right) \Big] dx &= 0, \quad m \in \mathbb{Z} \setminus \{0\},\label{6a} \\
    q_t - cq_x + \frac12 q_x^2 +g\eta - \frac{1}{2}\frac{\left(\eta_t-c\eta_x +\eta_xq_x\right)^2}{1+\eta_x^2} &= 0. \label{6b}
\end{align}
\end{subequations}
Unless otherwise stated, $x$ represents the horizontal coordinate in the traveling frame for the remainder of this work.

Non-dimensionalizing \eqref{6a}-\eqref{6b} according to $x \rightarrow x/\kappa$, $t \rightarrow t/\sqrt{g\kappa}$, $\eta \rightarrow \eta/\kappa$, $q \rightarrow q\sqrt{g/\kappa^3}$, $c \rightarrow c\sqrt{g/\kappa}$, and $h \rightarrow \alpha/\kappa$, we arrive at
\begin{subequations}
\begin{align}
    \int_{-\pi}^{\pi} e^{-i m x}\Big[\left(\eta_t-c\eta_x\right)\cosh\left( m\left(\eta + \alpha\right) \right) +iq_x\sinh\left(m\left(\eta + \alpha \right) \right) \Big] dx &= 0, \quad m \in \mathbb{Z} \setminus \{0\},\label{7a} \\
    q_t - cq_x + \frac12 q_x^2 +\eta - \frac{1}{2}\frac{\left(\eta_t-c\eta_x +\eta_xq_x\right)^2}{1+\eta_x^2} &= 0, \label{7b}
\end{align}
\end{subequations}
where $\alpha = \kappa h >0$ is the aspect ratio of the surface profile $\eta$ (in dimensional variables). Without loss of generality, we study solutions of the nondimensional equations \eqref{7a}-\eqref{7b}.

\vspace*{0.1in}

\noindent \textbf{Remark 1}. Dividing \eqref{7a} by $\cosh( m \alpha )$ and taking the limit $\alpha \rightarrow \infty$ yields (after some manipulation) the nonlocal equation in infinite depth:
\begin{align}
    \int_{-\pi}^{\pi} e^{-i m x +|m|\eta} \Big[\eta_t-c\eta_x + i \textrm{sgn}\left( m\right) q_x \Big] dx = 0, \quad m \in \mathbb{Z} \setminus \{0\}.
\end{align}
The local equation remains unchanged in infinite depth. 

\section{Small-Amplitude Stokes Waves}

Using the nondimensional AFM formulation \eqref{7a}-\eqref{7b}, Stokes waves are defined as surface displacements $\eta_S$ and velocity potentials (at the surface) $q_S$ that satisfy the following:
\begin{enumerate}
    \item[(i)] $\eta_S$ and $q_S$ are time-independent, infinitely smooth solutions of \eqref{7a}-\eqref{7b}. 
    \item[(ii)] $\eta_S$ and $q_{S,x}$ are $2\pi$-periodic with respect to $x$ (but not so of $q_S$). 
    \item[(iii)] $\eta_S$, $q_{S,x}$, and $c$ (the velocity of the Stokes wave) depend analytically on a small parameter $\varepsilon$ such that \begin{align} \eta_S\big|_{\varepsilon =0} = 0 = q_{S,x}\big|_{\varepsilon =0}  \quad \textrm{and} \quad ||\eta_S||_{\textrm{L}^2} = \varepsilon + \mathcal{O}\left(\varepsilon^2\right) \quad \textrm{as} \quad \varepsilon \rightarrow 0. \nonumber \end{align} 
    \item[(iv)] $\eta_S$ and $q_{S,x}$ are even in $x$ without loss of generality, and $c(\varepsilon)$ is even in $\varepsilon$.  
    \item[(v)] $\eta_S$ has zero average over one period.
\end{enumerate}
As mentioned in the Introduction, the existence of these waves is proven in \cite{levicivita25,nekrasov21,struik26}. In this section, we derive power series expansions of $\eta_S$, $q_{S,x}$, and $c$ in the small parameter $\varepsilon$ using the nondimensional AFM formulation. These expansions are required for the stability calculations considered in Sections 5 and 6.

Equating time derivatives to zero in \eqref{7a}-\eqref{7b} by property (i), integrating the $\cosh$ term in \eqref{7a} by parts using property (ii), and solving for $q_x$ in \eqref{7b}, we arrive at the following equations determining the Stokes waves:
\begin{subequations}
\begin{align}
\int_{-\pi}^{\pi} e^{-imx}\sqrt{\left(1+\eta_{S,x}^2\right)\left(c^2 - 2\eta_S \right)}&\sinh(m(\eta_S+\alpha)) dx = 0, \quad m \in \mathbb{Z} \setminus \{0\}, \label{8a} \\
q_{S,x} = c \pm& \sqrt{\left(1+\eta_{S,x}^2\right)\left(c^2 - 2\eta_S \right)}.  \label{8b}
\end{align}
\end{subequations}
By property (iii), the positive branch of \eqref{8b} is defined for left-traveling waves ($c<0$), while the negative branch is defined for right-traveling waves ($c>0$) \cite{constantinstrauss10}. In what follows, we consider right-traveling waves. Similar results hold for the other case.



\vspace*{0.1in}

\noindent \textbf{Remark 2}. In infinite depth, \eqref{8a} becomes 
\begin{align}
    \int_{-\pi}^{\pi} e^{-imx+|m|\eta_S}\sqrt{\left(1+\eta_{S,x}^2\right)\left(c^2 - 2\eta_S \right)}  dx = 0, \quad m \in \mathbb{Z} \setminus \{0\}.
\end{align}

\indent By properties (ii) and (iv), $\eta_S$ has a Fourier cosine series. We define the small-amplitude parameter $\varepsilon$ as the first Fourier cosine mode of $\eta_S$:
\begin{align}
    \varepsilon=\frac{1}{\pi}\int_{-\pi}^{\pi} \eta_S \cos(x) dx. \label{13}
\end{align}
Then, by property (iii),
\begin{align}
    \eta_S(x;\varepsilon) = \varepsilon \cos(x) + \mathcal{O}\left(\varepsilon^2\right),
\end{align}
for $|\varepsilon|\ll 1$. The leading-order term of $\eta_S$ completely resolves the first Fourier cosine mode: higher-order corrections do not include terms proportional to $\cos(x)$ as a result.

Using properties (iii) and (iv), we write $\eta_S$ and $c$ as power series in $\varepsilon$:
\begin{align}
    \eta_S(x;\varepsilon) &= \sum_{j=1}^{\infty} \eta_j(x)\varepsilon^j, \label{8.5a} \\
    c(\varepsilon) &= \sum_{j=0}^{\infty} c_{2j}\varepsilon^{2j}. \label{8.5b}
\end{align}
Both of these series are substituted into \eqref{8a} and, after equating powers of $\varepsilon$, a triangular sequence of linear integral equations for $\eta_j$(x) and $c_{2j}$ is found. Each of these integral equations depends on $m$, which can be any nonzero integer. 

\vspace*{0.1in}

\noindent \textbf{Remark 3}. Since $\eta_S$ is even in $x$, the integrand of \eqref{8a} modulo the complex exponential is even in $x$. Therefore, $m\in \mathbb{Z}^+$ without loss of generality.

\vspace*{0.1in}

The first nontrivial integral equation in this sequence is
\begin{align}
    \int_{-\pi}^{\pi}e^{-imx}\Big[mc_0^2\cosh(m\alpha)-\sinh(m\alpha) \Big]\eta_1(x) dx &= 0. \label{9}
\end{align}
From above, $\eta_1(x) = \cos(x)$. If \eqref{9} holds for all $m \in \mathbb{Z}^+$,
\begin{align}
    c_0^2 = \tanh(\alpha),
\end{align}
otherwise \eqref{9} is not satisfied when $m=1$. Since we study right-traveling waves, we choose $c_0>0$.

For the $j^{\textrm{th}}$ integral equation in the sequence ($j \geq 2$), one finds
\begin{subequations}
\begin{align}
    \eta_{j}(x) &= \sum_{\substack{\ell = 2 \\\ell~\textrm{even}}}^{j}\hat{N}_{j,\ell}\cos(\ell x)  \quad \textrm{for} \quad j ~\textrm{even}, \label{10a} \\
    \eta_{j}(x) &= \sum_{\substack{\ell = 3 \\\ell~\textrm{odd}}}^{j}\hat{N}_{j,\ell}\cos(\ell x) \quad \textrm{for} \quad j ~\textrm{odd}, \label{10b}
\end{align}
\end{subequations}
where the coefficients $\hat{N}_{j,\ell}$ are determined by the $j^{\textrm{th}}$ equation with $m = \ell$. No corrections to the velocity $c$ are found when $j$ is even. When $j$ is odd, $c_{j-1}$ is determined by the $j^{\textrm{th}}$ equation with $m = 1$, similar to the $j=1$ case considered above. This correction is chosen so that $\eta_j(x)$ has no terms proportional to $\cos(x)$. 



Expansions of $\eta_S$ and $c$ are substituted into \eqref{8b}. After equating powers of $\varepsilon$, an expansion for $q_{S,x}$ follows immediately. In general,
\begin{align}
    q_{S,x}(x;\varepsilon) = \sum_{j=1}^{\infty} q_{j,x}(x)\varepsilon^j. \label{11}
\end{align}
The corrections $q_{j,x}(x)$ have the same structure as \eqref{10a}-\eqref{10b}, but also include constant modes (when $j$ is even) and modes proportional to $\cos(x)$ (when $j$ is odd). Thus, $q_{S,x}$ has nonzero average, and the first Fourier cosine mode of $q_{S,x}$ has corrections beyond $\mathcal{O}\left(\varepsilon\right)$, unlike $\eta_S$.

\vspace*{0.1in}
\noindent \textbf{Remark 4}. Integrating \eqref{11} term-by-term gives $q_S$. The constant of integration can be eliminated by a Galilean transformation of \eqref{8b}. Because $q_{S,x}$ has nonzero average, $q_S$ exhibits linear growth in $x$. This behavior captures the mean flow induced by the traveling frame.

\vspace*{0.1in}

Explicit representations for the expansions of $\eta_S$, $q_{S,x}$, and $c$ up to $\mathcal{O}\left(\varepsilon^4\right)$ are found in Appendix A. In Figure \ref{fig4}, these expansions show excellent agreement with direct numerical computations of the Stokes waves using the continuation method presented in \cite{deconinckoliveras11}. 
 
 \begin{figure}[tb]
    \hspace*{-0.4cm}
    \includegraphics[height=7.5cm,width=16cm]{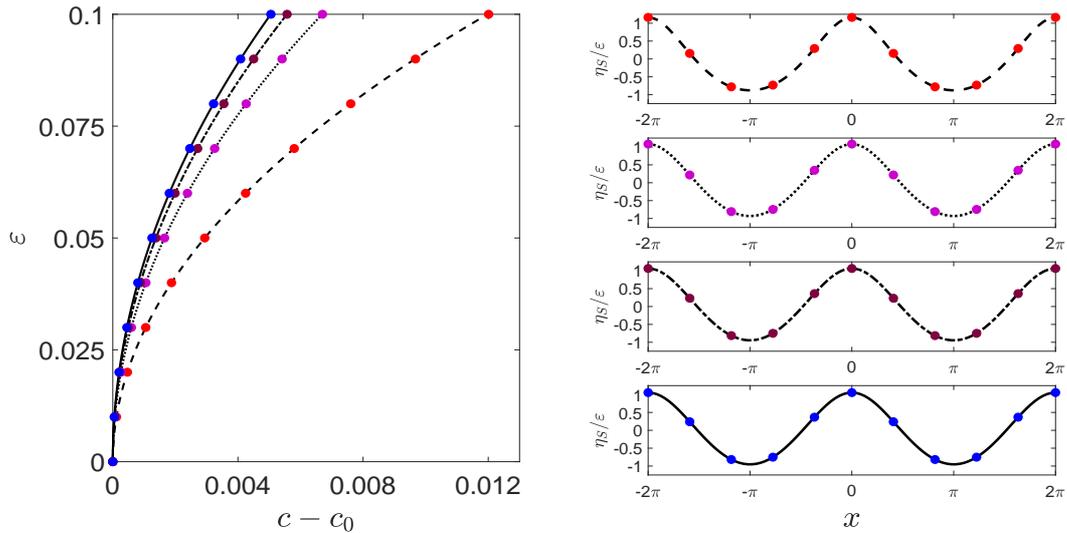}
    \caption{(Left) The amplitude \emph{vs}. velocity bifurcation diagram of $2\pi$-periodic Stokes waves when $\alpha = 1$ (dashed line), $\alpha = 1.5$ (dotted line), $\alpha = 2$ (dot-dashed line), and $\alpha = \infty$ (solid line), according to our $\mathcal{O}\left(\varepsilon^4\right)$ asymptotic calculations. The zeroth-order contribution $c_0$ is removed for better visibility. The numerical results are given by the colored dots. Red dots correspond to $\alpha = 1$, magenta dots correspond to $\alpha = 1.5$, purple dots correspond to $\alpha = 2$, and blue dots correspond to $\alpha = \infty$. (Right) Expansions of $\eta_S/\varepsilon$ to $\mathcal{O}\left(\varepsilon^4\right)$ with $\varepsilon = 0.1$ for $\alpha = 1, 1.5, 2,$ and $\infty$ (arranged from top to bottom using the same line styles as in the left figure). A sampling of the numerical results is given by the colored dots using the same color scheme as in the left figure.\\}
    \label{fig4}
\end{figure} 

\section{The Spectral Instability of Stokes Waves}

\subsection{The Stability Spectrum}
We consider perturbations to the Stokes waves of the form
\begin{align}
    \begin{pmatrix} \eta(x,t;\varepsilon,\rho) \\ q(x,t;\varepsilon,\rho) \end{pmatrix} = \begin{pmatrix} \eta_S(x;\varepsilon) \\ q_S(x;\varepsilon) \end{pmatrix} + \rho \begin{pmatrix} \eta_\rho(x,t; \varepsilon) \\ q_\rho(x,t; \varepsilon) \end{pmatrix} + \mathcal{O}\left(\rho^2\right), \label{14}
\end{align}
where $|\rho|\ll 1$ is a parameter independent of $\varepsilon$. The perturbations $\eta_\rho$ and $q_\rho$ are sufficiently smooth functions of $x$ and $t$ that are bounded over the real line for each $t\geq 0$.

The nonlocal equation \eqref{7a} assumes $\eta$, $\eta_t$, and $q_x$ are $2\pi$-periodic in $x$, which is not required of our perturbations. We modify \eqref{7a} to allow $\eta, \eta_t,$ and $q_x \in C^0(\mathbb{R}) \cap L^{\infty}(\mathbb{R})$ for each $t \geq 0$. The appropriate modification \cite{deconinckoliveras11} is
\begin{align}
    \left<  e^{-i k x}\Big[\left(\eta_t-c\eta_x\right)\cosh\left(k\left(\eta + \alpha \right) \right) +iq_x\sinh\left(k\left(\eta + \alpha \right) \right) \Big] \right> = 0, \quad k \in \mathbb{R} \setminus \{0\}, \label{5a} 
\end{align}
where
\begin{align}
    \left< f(x) \right> = \displaystyle \lim_{L \rightarrow \infty} \frac{1}{L} \int_{-L/2}^{L/2} f(x) dx,
\end{align}
for any $f(x) \in C^0(\mathbb{R}) \cap L^{\infty}(\mathbb{R})$ \cite{bohr47,deconinckoliveras11}. If $\eta, \eta_t,$ and $q_x$ are $2\pi$-periodic in $x$ for each $t\geq0$, then \eqref{5a} reduces to \eqref{7a}.

Substituting \eqref{14} into \eqref{7b} and \eqref{5a} and equating powers of $\rho$, terms of $\mathcal{O}\left(\rho^0\right)$ necessarily cancel, since $\eta_S$ and $q_S$ solve \eqref{7b} and \eqref{5a}. At $\mathcal{O}\left(\rho\right)$, one finds the governing equations for $\eta_\rho$ and $q_\rho$:
\begin{subequations}
\begin{align}
   \left<e^{-ikx}\big[c\mathcal{C}_k\eta_{\rho,x} +k\left(c\mathcal{S}_k\eta_{S,x} -i\mathcal{C}_kq_{S,x} \right)\eta_\rho -i\mathcal{S}_kq_{\rho,x}\big] \right>  &= \left< e^{-ikx}\mathcal{C}_k\eta_{\rho,t} \right>, \label{15} \\ 
    \eta_{S,x}{\zeta}^2\eta_{\rho,x} - \eta_{\rho} -{\zeta}q_{\rho,x}  &=  q_{\rho,t} -\eta_{S,x}{\zeta}\eta_{\rho,t}, \label{16}
\end{align}
\end{subequations}
where 
\begin{align} \mathcal{C}_k = \cosh(k(\eta_S+\alpha)), \quad \quad \mathcal{S}_k = \sinh(k(\eta_S+\alpha)), 
\quad \quad {\zeta} =\frac{q_{S,x}-  c}{1+\eta_{S,x}^2}.
\end{align} 

Equations \eqref{15}-\eqref{16} are autonomous in $t$. We separate variables to find
\begin{align}
    \begin{pmatrix} 
    \eta_\rho(x,t) \\
    q_\rho(x,t) 
    \end{pmatrix} = e^{\lambda t}\begin{pmatrix} N(x) \\ Q(x) \end{pmatrix},
\end{align}
where $\lambda \in \mathbb{C}$ controls the growth rates of the perturbations. The functions $N(x)$ and $Q(x)$ satisfy 
\begin{subequations}
\begin{align}
\left<e^{-ikx}\big[c\mathcal{C}_kN_x +k\left(c\mathcal{S}_k\eta_{S,x} -i\mathcal{C}_kq_{S,x} \right)N -i\mathcal{S}_kQ_x\big] \right>     &= \lambda \left< e^{-ikx}\mathcal{C}_kN \right>, \label{17}\\
    \eta_{S,x}{\zeta}^2N_x - N -{\zeta}Q_x &=   \lambda\left(Q -\eta_{S,x}{\zeta}N\right). \label{18}
\end{align}
\end{subequations}
Equations \eqref{17}-\eqref{18} are invariant under the shift $x \rightarrow x + 2\pi$ by the periodicity of $\eta_S$ and $q_{S,x}$. Therefore, we expect the solutions $N$ and $Q$ to have Bloch form \cite{deconinckoliveras11}
\begin{align}
    \begin{pmatrix} N(x) \\ Q(x) \end{pmatrix} = e^{i\mu x}\begin{pmatrix} \mathcal{N}(x) \\ 
    \mathcal{Q}(x) \end{pmatrix}, \label{19}
\end{align}
where $\mu \in \mathbb{R}$ is the Floquet exponent and $\mathcal{N}$ and $\mathcal{Q}$ are sufficiently smooth and $2\pi$-periodic. 
Note that by redefining $\mathcal{N}$ and $\mathcal{Q}$, $\mu \in [-1/2,1/2)$, without loss of generality.

Substituting \eqref{19} into \eqref{17}-\eqref{18}, we arrive at 
\begin{subequations}
\begin{align}
\left<e^{-i(k-\mu)x}\big[c\mathcal{C}_k\mathcal{D}_x\mathcal{N} +k\left(c\mathcal{S}_k\eta_{S,x} -i\mathcal{C}_kq_{S,x} \right)\mathcal{N} -i\mathcal{S}_k\mathcal{D}_x\mathcal{Q} \big] \right>    &=   \lambda \left< e^{-i(k-\mu)x}\mathcal{C}_k\mathcal{N} \right>, \label{20} \\
    \eta_{S,x}{\zeta}^2\mathcal{D}_x\mathcal{N} - \mathcal{N} -{\zeta}\mathcal{D}_x\mathcal{Q} &=  \lambda\left(\mathcal{Q} -\eta_{S,x}{\zeta}\mathcal{N}\right) ,\label{21}
\end{align}
\end{subequations}
where $\mathcal{D}_x = i\mu + \partial_x$.

The integrands of the averaging operators in \eqref{20} are $2\pi$-periodic except for the complex exponentials. These operators evaluate to zero unless $k-\mu = n \in \mathbb{Z}$ \cite{deconinckoliveras11}. For such $k$, \eqref{20} becomes
\begin{align} 
   \left<e^{-inx}\big[c\mathcal{C}_{n+\mu}\mathcal{D}_x\mathcal{N} +(n+\mu)\left(c\mathcal{S}_{n+\mu}\eta_{S,x} -i\mathcal{C}_{n+\mu}q_{S,x} \right)\mathcal{N} -i\mathcal{S}_{n+\mu}\mathcal{D}_x\mathcal{Q} \big] \right> = \lambda \left<e^{-inx}\mathcal{C}_{n+\mu}\mathcal{N} \right>, \quad n \in \mathbb{Z}. \label{22}  
\end{align}
The averaging operators of \eqref{22} reduce to Fourier transforms:
\begin{align}
    \left<e^{-inx}f(x)\right> = \frac{1}{2\pi}\int_{-\pi}^{\pi} e^{-inx} f(x) dx = \mathcal{F}_n[f(x)],
\end{align}
for any $f(x) \in L^2_{\textrm{per}}\left(-\pi,\pi\right)$. The inverse transform is
\begin{align}
    \mathcal{F}^{-1}[\{f_n\}] = \sum_{n=-\infty}^{\infty} f_n e^{inx},
\end{align}
provided $\{f_n\} \in \ell^2(\mathbb{Z})$. Using the inverse transform on \eqref{22}, we find
\begin{align}
     \sum_{n=-\infty}^{\infty}e^{inx}\mathcal{F}_n\big[c\mathcal{C}_{n+\mu}\mathcal{D}_x\mathcal{N} +(n+\mu)\left(c\mathcal{S}_{n+\mu}\eta_{S,x} -i\mathcal{C}_{n+\mu}q_{S,x} \right)\mathcal{N} \big]\quad\quad& \label{23} \\ + \sum_{n=-\infty}^{\infty}e^{inx}\mathcal{F}_n\big[-i\mathcal{S}_{n+\mu}\mathcal{D}_x\mathcal{Q} \big]&=\lambda \sum_{n=-\infty}^{\infty}e^{inx}\mathcal{F}_n\left[\mathcal{C}_{n+\mu}\mathcal{N} \right]. \nonumber
\end{align}
Equations \eqref{21} and \eqref{23} are written compactly as
\begin{align}
    \mathcal{L}_{\mu,\varepsilon} {\bf w}_{\mu,\varepsilon} = \lambda_{\mu,\varepsilon} \mathcal{R}_{\mu,\varepsilon} {\bf w}_{\mu,\varepsilon},\label{24}
\end{align}
where $\lambda =\lambda_{\mu,\varepsilon}$, ${\bf w}_{\mu,\varepsilon} = (\mathcal{N},\mathcal{Q})^T$, and
\begin{align}
\mathcal{L}_{\mu,\varepsilon} = \begin{pmatrix} \mathcal{L}_{\mu,\varepsilon}^{(1,1)}& \mathcal{L}_{\mu,\varepsilon}^{(1,2)} \\ \mathcal{L}_{\mu,\varepsilon}^{(2,1)} & \mathcal{L}_{\mu,\varepsilon}^{(2,2)} \end{pmatrix}, \quad  \mathcal{R}_{\mu,\varepsilon} = \begin{pmatrix} \mathcal{R}_{\mu,\varepsilon}^{(1,1)}& 0 \\ \mathcal{R}_{\mu,\varepsilon}^{(2,1)} & 1 \end{pmatrix},
\end{align} \vspace*{-0.4cm}
\begin{subequations}
\allowdisplaybreaks
\begin{flalign}
    \mathcal{L}_{\mu,\varepsilon}^{(1,1)}[\mathcal{N}] &=  \sum_{n=-\infty}^{\infty}e^{inx} \mathcal{F}_n \big[c\mathcal{C}_{n+\mu}\mathcal{D}_x\mathcal{N} +(n+\mu)\left(c\mathcal{S}_{n+\mu}\eta_{S,x} -i\mathcal{C}_{n+\mu}q_{S,x} \right)\mathcal{N}\big], \\
   \mathcal{L}_{\mu,\varepsilon}^{(1,2)}[\mathcal{Q}]  &=  \sum_{n=-\infty}^{\infty}e^{inx}\mathcal{F}_n\big[-i\mathcal{S}_{n+\mu}\mathcal{D}_x\mathcal{Q} \big],\\
    \mathcal{L}_{\mu,\varepsilon}^{(2,1)}[\mathcal{N}] &= \eta_{S,x}{\zeta}^2\mathcal{D}_x\mathcal{N} - \mathcal{N}, \\
     \mathcal{L}_{\mu,\varepsilon}^{(2,2)}[\mathcal{Q}] &= -{\zeta}\mathcal{D}_x\mathcal{Q}, \\
     \mathcal{R}_{\mu,\varepsilon}^{(1,1)}[\mathcal{N}] &= \sum_{n=-\infty}^{\infty}e^{inx}\mathcal{F}_n\left[\mathcal{C}_{n+\mu}\mathcal{N} \right], \\ \mathcal{R}_{\mu,\varepsilon}^{(2,1)}[\mathcal{N}] &= -\eta_{S,x}{\zeta}\mathcal{N}.
\end{flalign}
\end{subequations}
Equation \eqref{24} represents a two-parameter family of generalized eigenvalue problems for the linear operators $\mathcal{L}_{\mu,\varepsilon}$ and $\mathcal{R}_{\mu,\varepsilon}$. 
 \vspace*{0.1in}
 
 \noindent \textbf{Remark 5}. In infinite depth, 
 \begin{subequations}
 \begin{flalign}
    \mathcal{L}_{\mu,\varepsilon}^{(1,1)}[\mathcal{N}] &= \sum_{n=-\infty}^{\infty} e^{inx}\mathcal{F}_n\big[e^{|n+\mu|\eta_S}\big(c\mathcal{D}_x\mathcal{N} +\big(c\eta_{S,x}|n+\mu|-i(n+\mu)q_{S,x}\big)\mathcal{N} \big) \big], \\
    \mathcal{L}_{\mu,\varepsilon}^{(1,2)}[\mathcal{Q}] &= \sum_{n=-\infty}^{\infty} e^{inx}\mathcal{F}_n\big[e^{|n+\mu| \eta_S}\big(-i\textrm{sgn}(n+\mu)\mathcal{D}_x\mathcal{Q} \big)\big], \\
    \mathcal{R}_{\mu,\varepsilon}^{(1,1)}[\mathcal{N}] &= \sum_{n=-\infty}^{\infty} e^{inx}\mathcal{F}_n\big[e^{|n+\mu|\eta_S}\mathcal{N}\big].
 \end{flalign}
 \end{subequations}
 All other entries are the same as above.

\vspace*{0.1in}

 The spectrum of \eqref{24} has a countable collection of finite-multiplicity eigenvalues $\lambda_{\mu,\varepsilon}$ for each $\mu$ \cite{akersnicholls14,deconinckoliveras11,kapitulapromislow13}. The union of these eigenvalues over $\mu \in [-1/2,1/2)$ is defined as the stability spectrum of Stokes waves with amplitude $\varepsilon$. If there exists $\lambda_{\mu,\varepsilon}$ for some $\mu$ such that $\Re\left(\lambda_{\mu,\varepsilon} \right)>0$, then there exist perturbations of the Stokes waves $\eta_{\rho}$ and $q_{\rho}$ that grow exponentially in time. In this case, the Stokes waves are spectrally unstable. If no such $\mu$ and $\lambda_{\mu,\varepsilon}$ exist, the Stokes waves are spectrally stable.
 
 

\subsection{Necessary Conditions for High-Frequency Instabilities}
When $\varepsilon=0$, \eqref{24} reduces to a generalized eigenvalue problem with constant coefficients:
\begin{align} {\small
    \begin{pmatrix} ic_0(\mu+D)\cosh(\alpha(\mu+D)) & (\mu+D)\sinh(\alpha(\mu+D)) \\ -1 & ic_0(\mu+D) \end{pmatrix} {\bf w}_{\mu,0} = \lambda_{\mu,0}\begin{pmatrix} \cosh(\alpha(\mu+D)) & 0 \\ 0 & 1 \end{pmatrix}{\bf w}_{\mu,0},} \label{25}
\end{align}
where $D=-i\partial_x$. The eigenvalues of \eqref{25} are
\begin{align}
   \lambda_{\mu,0,n}^{(\sigma)} = -i\Omega_{\sigma}(\mu+n), \quad \quad \sigma = \pm 1, \quad \quad n \in \mathbb{Z},\label{26}
\end{align}
with \begin{subequations}
\begin{align}
    \Omega_{\sigma}(z) &=-c_0z +\sigma \omega(z), \label{26a}\\ \omega(z) &= \textrm{sgn}(z)\sqrt{z\tanh(\alpha z)}. \label{26b}
\end{align}
\end{subequations}
Equation \eqref{26a} is the linear dispersion relation of the nondimensional Euler equations in a frame traveling with velocity $c_0$. The parameter $\sigma$ specifies the branch of the dispersion relation. As expected, \eqref{26} gives a countable collection of eigenvalues for each $\mu \in [-1/2,1/2)$. These eigenvalues are purely imaginary, and therefore, the zero-amplitude Stokes waves are spectrally stable.

High-frequency instabilities develop from nonzero eigenvalues of \eqref{25} that have double (algebraic and geometric) multiplicity for a Floquet exponent $\mu_0$ that satisfies \cite{akersnicholls12,deconincktrichtchenko17,mackaysaffman86}: 
\begin{align}
    \lambda^{(\sigma_1)}_{\mu_0,0,n} = \lambda^{(\sigma_2)}_{\mu_0,0,n+p} \neq 0,\label{27}
\end{align}
for $p \in \mathbb{Z} \setminus \{0\}$. Such double eigenvalues occur only if $\sigma_1 \neq \sigma_2$ and $|p|>1$ \cite{deconincktrichtchenko17}. More specifically, we have the following theorem: 

\vspace*{0.1in}

\noindent \textbf{Theorem 1}. Let $c_0>0$, $\sigma_1 = 1$, and $\sigma_2 = -1$. For each $p \in \mathbb{Z} \setminus \{0,\pm 1\}$, there exists a unique Floquet exponent $\mu_{0,p} \in [-1/2,1/2)$ and unique integer $n_p$ such that 
\begin{align}
    \lambda_{0,p} = \lambda^{(1)}_{\mu_{0,p},0,n_p} = \lambda^{(-1)}_{\mu_{0,p},0,n_p+p} \label{28} \neq 0.
\end{align}
The eigenvalues have the symmetry $\lambda_{0,-p} = -\lambda_{0,p}$, and the magnitudes of the eigenvalues are strictly monotonically increasing as $|p| \rightarrow \infty$. The corresponding eigenfunctions are
\begin{align}
    {\bf w}_{0,p} = \beta_0 \begin{pmatrix} 1 \\ \frac{-i}{\omega(n_p+\mu_{0,p})} \end{pmatrix} e^{in_px} + \gamma_0 \begin{pmatrix} 1 \\ \frac{i}{\omega(n_p+p+\mu_{0,p})} \end{pmatrix} e^{i(n_p+p)x}, \label{28.5}
\end{align}
where $\omega$ is given by \eqref{26b} and $\beta_0,\gamma_0 \in \mathbb{C} \setminus \{0\}$. 

\vspace*{0.1in}

An important corollary is the following:

\vspace*{0.1in}

\noindent \textbf{Corollary 1}. Let $c_0>0$. Let $\lambda_{0,p}$ be given by \eqref{28} for some $p \in \mathbb{Z} \setminus \{0,\pm 1\}$. Then,
\begin{align}
    \omega(n_p+\mu_{0,p})\omega(n_p+p+\mu_{0,p})>0, \label{29}
\end{align}
and
\begin{align}
    c_{g,1}(n_p+\mu_{0,p}) \neq c_{g,-1}(n_p+p+\mu_{0,p}), \label{30}
\end{align}
where $c_{g,\sigma}(z)$ is the group velocity of $\Omega_{\sigma}(z)$, \emph{i.e.}, $c_{g,\sigma}(z) = \Omega_{\sigma,z}(z).$

\vspace*{0.1in}

Similar results hold if $c_0<0$ provided $\sigma_1 = -1$ and $\sigma_2 = 1$. See \cite{creedonetal21b} for the proofs of Theorem 1 and Corollary 1. 

The product \eqref{29} is equivalent to the Krein condition developed by MacKay \emph{\&} Saffman (1986) \cite{mackaysaffman86} and, in more generality, Deconinck \emph{\&} Trichtchenko (2017) \cite{deconincktrichtchenko17}. This is a second necessary condition for high-frequency instabilities. Corollary 1 guarantees this condition is satisfied for all nonzero eigenvalues of \eqref{25} with double multiplicity. Both \eqref{29} and \eqref{30} are crucial to the formal asymptotic expansions of the high-frequency instabilities derived in Sections 5 and 6.

\vspace*{0.1in}

\noindent \textbf{Remark 6}. In infinite depth, $\mu_{0,p}$ and $\lambda_{0,p}$ are known explicitly. For $c_0>0$, 
\begin{subequations}
\begin{align}
    \mu_{0,p} &= -\frac{\textrm{sgn}(p)}{8}\big((-1)^p+1\big),\\
    \lambda_{0,p} &= i\frac{\textrm{sgn}(p)}{4}\big(1-p^2\big).
\end{align}
\end{subequations}
These eigenvalues have the conjugate symmetry $\lambda_{0,-p} = -\lambda_{0,p}$, and $\{|\lambda_{0,p}|\}$ is strictly monotonically increasing as $|p| \rightarrow \infty$, similar to the finite-depth case.

\section{First Isola. High-Frequency Instabilities: $p=2$}

We develop a perturbation method to obtain the leading-order behavior of the high-frequency isola that arises from $\lambda_{0,p}$ with $p=2$. According to Theorem 1, this isola is the closest to the origin. We assume the spectral data of \eqref{24} corresponding to the isola vary analytically with $\varepsilon$, including the Floquet exponent:
\begin{subequations}
\begin{align}
\lambda_{\mu(\varepsilon),\varepsilon} &= \lambda_{0,p} + \lambda_1 \varepsilon+ \lambda_2 \varepsilon^2 +\mathcal{O}\left(\varepsilon^3\right), \label{29a} \\
{\bf w}_{\mu(\varepsilon),\varepsilon} &= {\bf w}_{0,p} + {\bf w}_1 \varepsilon + {\bf w}_2 \varepsilon^2 + \mathcal{O}\left(\varepsilon^3\right), \\
\mu(\varepsilon) &= \mu_{0,p} + \mu_1\varepsilon + \mu_2\varepsilon^2 + \mathcal{O}\left(\varepsilon^3\right). \label{29c}
\end{align}
\end{subequations}
If the Floquet exponent has no dependence on $\varepsilon$, the expansions above are justified by standard eigenvalue perturbation theory \cite{kato66}, and one can find at most two eigenvalues on the isola. In contrast, by expanding the Floquet exponent as a series in $\varepsilon$, we asymptotically approximate all the eigenvalues on the isola for sufficiently small $\varepsilon$. We see below that the leading-order behavior of these eigenvalues is obtained at $\mathcal{O}\left(\varepsilon^2\right)$.

\vspace*{0.1in}

\noindent \textbf{Remark 7}. Choosing $p=-2$ gives the isola conjugate to the $p=2$ isola. Thus, we choose $p=2$ without loss of generality. 

\vspace*{0.1in}

We impose the following normalization on ${\bf w}_{\mu(\varepsilon),\varepsilon}$: 
\begin{align}
    \mathcal{F}_{n_p}[{\bf w}_{\mu(\varepsilon),\varepsilon}\cdot{\bf e}_1] = 1, \label{28.75}
\end{align}
where $n_p \in \mathbb{Z}$ is given by Theorem 1 and ${\bf e}_1 = (1,0)^T$. Then, $\beta_0 = 1$ in \eqref{28.5}, and all subsequent corrections of ${\bf w}_{\mu(\varepsilon),\varepsilon}$ do not include the Fourier mode $\textrm{exp}(in_px)$ in the first component. The eigenvalue and Floquet expansions, \eqref{29a} and \eqref{29c} above, are unaffected by this normalization. For ease of notation, let $\lambda_{0,p} \rightarrow \lambda_0$, ${\bf w}_{0,p} \rightarrow {\bf w}_0$, $\mu_{0,p} \rightarrow \mu_0$, and $n_p \rightarrow n$. 

Several of the asymptotic expressions that follow are suppressed for ease of readability. See the Data Availability Statement at the end of this manuscript for access to the full expressions.
\subsection{The $\mathcal{O}\left(\varepsilon\right)$ Problem}
Substituting expansions \eqref{29a}-\eqref{29c} into the generalized eigenvalue problem \eqref{24} and equating powers of $\varepsilon$, terms of $\mathcal{O}\left(\varepsilon^0\right)$ cancel by the choice of $\lambda_{0}$, ${\bf w}_0$, and $\mu_0$. Terms of $\mathcal{O}\left(\varepsilon\right)$ yield
\begin{align}
\left(L_0-\lambda_{0}R_0\right){\bf w}_1 = \left(\lambda_1R_0 - \left(L_1 - \lambda_{0} R_1 \right) \right){\bf w}_0,\label{30}
\end{align}
where
\begin{align}
    L_j = \frac{1}{j!} \frac{\partial^j\mathcal{L}_{\mu(\varepsilon),\varepsilon}}{\partial \varepsilon^j}\Big|_{\varepsilon = 0}, \quad \quad     R_j = \frac{1}{j!} \frac{\partial^j\mathcal{R}_{\mu(\varepsilon),\varepsilon}}{\partial \varepsilon^j}\Big|_{\varepsilon = 0}, \quad \quad j \in \mathbb{W}.
\end{align}

If \eqref{30} can be solved for ${\bf w}_1$, the inhomogeneous terms on the RHS of \eqref{30} must be orthogonal to the nullspace of the adjoint of $L_0-\lambda_{0}R_0$ by the Fredholm alternative. A direct calculation shows
\begin{align}
    \textrm{Null}\left(\left(L_0-\lambda_{0}R_0\right)^\dagger\right) = \textrm{Span}\left\{ \begin{pmatrix} 1 \\ -i\omega\left(n+\mu_0\right) \end{pmatrix} e^{in x}, \begin{pmatrix} 1 \\ i\omega\left(n+p+\mu_0\right) \end{pmatrix} e^{i(n+p)x} \right\}.
\end{align}
Hence, we impose the following solvability conditions on \eqref{30}: 
\begin{subequations}
\begin{align}
    \left< \begin{pmatrix} 1 \\ -i\omega\left(n+\mu_0\right) \end{pmatrix} e^{in x},\left(\lambda_1R_0 - \left(L_1 - \lambda_{0} R_1 \right) \right){\bf w}_0\right> &= 0, \\
    \left<\begin{pmatrix} 1 \\ i\omega\left(n+p+\mu_0\right) \end{pmatrix} e^{i(n+p)x} ,\left(\lambda_1R_0 - \left(L_1 - \lambda_{0} R_1 \right) \right){\bf w}_0 \right> &= 0, 
\end{align}
\end{subequations}
where $\left<\cdot,\cdot\right>$ is the standard inner-product on $\textrm{L}^2_{\textrm{per}}(-\pi,\pi)\times \textrm{L}^2_{\textrm{per}}(-\pi,\pi)$. Simplifying both conditions, we arrive at 
\begin{subequations}
\begin{align}
    \lambda_1 + i\mu_1c_{g,1}\left(n+\mu_0\right) &= 0,\\
    \gamma_0\left(\lambda_1 + i\mu_1c_{g,-1}\left(n+p+\mu_0 \right)\right) &= 0.
\end{align}
\end{subequations}
Since $\gamma_0 \neq 0$ by Theorem 1 and $c_{g,1}\left(n+\mu_0\right) \neq c_{g,-1}\left(n+p+\mu_0 \right)$ by Corollary 1, we must have
\begin{align}
    \lambda_1 = 0 = \mu_1. \label{31}
\end{align}
Thus no instabilities are found at $\mathcal{O}\left(\varepsilon\right)$. 

Before proceeding to $\mathcal{O}\left(\varepsilon^2\right)$, we invert $L_0-\lambda_{0}R_0$ against its range to find the particular solution of ${\bf w}_1$. Uniting the particular solution with the nullspace of $L_0-\lambda_{0}R_0$, 
\begin{align}
    {\bf w}_1 = \sum_{\substack{j = n-1 \\ j \neq n, n+p}}^{n+p+1} \hat{\mathcal{W}}_{1,j}e^{ijx} + \beta_1 \begin{pmatrix} 1 \\ \frac{-i}{\omega(n+\mu_0)} \end{pmatrix} e^{inx} + \gamma_1 \begin{pmatrix} 1 \\ \frac{i}{\omega(n+p+\mu_0)} \end{pmatrix} e^{i(n+p)x},
\end{align}
where the coefficients $\hat{\mathcal{W}}_{1,j}$ depend on $\alpha$ (possibly through intermediate dependencies on known zeroth-order results) and at most linearly on $\gamma_0$. The parameter $\gamma_1 \in \mathbb{C}$ is free at this order. By our choice of normalization \eqref{28.75}, $\beta_1 = 0$. Thus,
\begin{align}
    {\bf w}_1 = \sum_{\substack{j = n-1 \\ j \neq n, n+p}}^{n+p+1} \hat{\mathcal{W}}_{1,j}e^{ijx} + \gamma_1 \begin{pmatrix} 1 \\ \frac{i}{\omega(n+p+\mu_0)} \end{pmatrix} e^{i(n+p)x}.
\end{align}

\subsection{The $\mathcal{O}\left(\varepsilon^2\right)$ Problem}

At $\mathcal{O}\left(\varepsilon^2\right)$, the spectral problem \eqref{24} is
\begin{align}
    \left(L_0 - \lambda_{0}R_0\right){\bf w}_2 = \lambda_2 R_0{\bf w}_0 - \left(L_1 - \lambda_{0}R_1 \right){\bf w}_1 - \left(L_2 - \lambda_{0}R_2\right){\bf w}_0, \label{32}
\end{align}
using \eqref{31}. Proceeding as above, we obtain the solvability conditions for \eqref{32}:
\begin{subequations}
\begin{align}
    2\left(\lambda_2 + i\mathfrak{c}_{2,1,n}\right) + i\gamma_0\mathfrak{s}_{2,n} &= 0, \label{32a} \\
    2\gamma_0\left(\lambda_2 + i\mathfrak{c}_{2,-1,n+p}\right) + i \mathfrak{s}_{2,n+p} &= 0, \label{32b}
\end{align}
\end{subequations}
where 
\begin{align}
    \mathfrak{c}_{2,\sigma,j} &= \mu_2c_{g,\sigma}\left(j+\mu_0\right) - \mathfrak{p}_{2,j}.
\end{align}
The quantities $\mathfrak{s}_{2,j}$ and $\mathfrak{p}_{2,j}$ depend only on $\alpha$ (possibly through known zeroth- and first-order quantities). Using the collision condition \eqref{27}, it can be shown that the product of $\mathfrak{s}_{2,n}$ and $\mathfrak{s}_{2,n+p}$ is related to a perfect square:
\begin{align}
    \mathfrak{s}_{2,n}  \mathfrak{s}_{2,n+p} &= - \frac{\mathcal{S}_2^2}{ \omega(n+\mu_0)\omega(n+p+\mu_0)},
\end{align}
where
\begin{align}
    \mathcal{S}_2 =&~ \mathcal{T}_{2,1} + \mathcal{T}_{2,2}\hat{N}_{2,2} + \mathcal{T}_{2,3}\hat{Q}_{2,2}. \nonumber
\end{align}
The expressions $\mathcal{T}_{2,j}$ are functions only of $\alpha$, as are the Stokes wave corrections $\hat{N}_{2,2}$ and $\hat{Q}_{2,2}$, see Appendix A. When fully expanded, $\mathcal{S}_2$ consists of roughly 100 terms (depending on how it is written), but each term depends only on $\alpha$. The full expression of $\mathcal{S}_2$ is found in the appropriate Mathematica notebook provided in the Data Availability Statement.

Solving for $\lambda_2$ in \eqref{32a}-\eqref{32b}, 
\begin{align}\label{33}
    \lambda_2 &= -i\biggr( \frac{\mathfrak{c}_{2,-1,n+p} + \mathfrak{c}_{2,1,n}}{2} \biggr) 
    \pm \sqrt{-\biggr(\frac{\mathfrak{c}_{2,-1,n+p} - \mathfrak{c}_{2,1,n}}{2} \biggr)^2 + \frac{\mathcal{S}_2^2}{4\omega(n+\mu_0)\omega(n+p+\mu_0)} }.
\end{align}

From Corollary 1, $\omega(n+\mu_0)\omega(n+p+\mu_0)>0$. Thus, $\lambda_2$ has nonzero real part for $\mu_2 \in \left(M_{2,-},M_{2,+}\right)$, where
\begin{align}
    M_{2,\pm} = \mu_{2,*} \pm \frac{\left|\mathcal{S}_2\right|}{\left|c_{g,-1}\left(n+p+\mu_0\right)-c_{g,1}\left(n+\mu_0\right) \right| \sqrt{\omega(n+\mu_0)\omega(n+p+\mu_0)}}, \label{34}
\end{align}
and
\begin{align}
    \mu_{2,*} = \frac{\mathfrak{p}_{2,n+p}-\mathfrak{p}_{2,n}}{c_{g,-1}\left(n+p+\mu_0\right)-c_{g,1}\left(n+\mu_0\right) }, \label{35}
\end{align}
\sloppypar \noindent provided $\mathcal{S}_2 \not\equiv 0$. Note that  Corollary 1 guarantees \eqref{34} and \eqref{35} are well-defined, since $c_{g,-1}\left(n+p+\mu_0\right)$ and $ c_{g,1}\left(n+\mu_0\right) $ are never equal. 

\begin{figure}[tb]
    \hspace*{-0.0cm}
    \includegraphics[height=7.5cm,width=15.5cm]{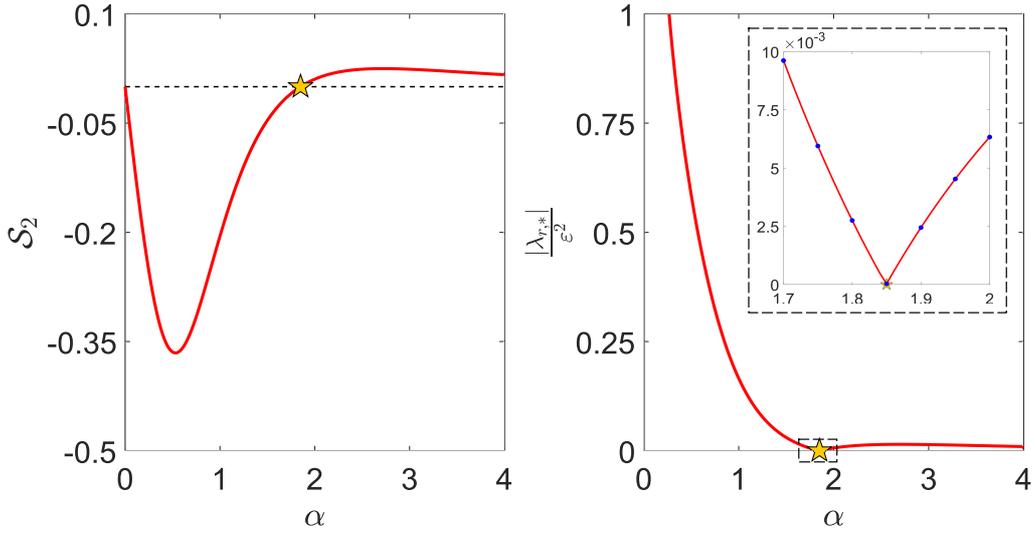}
    \caption{(Left) A plot of $\mathcal{S}_2$ \emph{vs.} $\alpha$ (solid red). The zero of $\mathcal{S}_2$ for $\alpha > 0$ is $\alpha_1 =  1.8494040837...$ (gold star). (Right) The real part $\lambda_{r,*}$ of the most unstable eigenvalue on the $p=2$ isola as a function of $\alpha$ according to our asymptotic calculations (solid red). The real part of the eigenvalue is normalized by $\varepsilon^2$ for better visibility. We zoom-in around $\alpha = \alpha_1$ (gold star) in the inlay. The real part of the most unstable eigenvalue on the isola vanishes as $\alpha \rightarrow \alpha_1$ according to our asymptotic calculations, which agrees with our numerical results using the FFH method with $\varepsilon = 0.01$ (blue dots).}
    \label{fig5}
\end{figure}

A plot of $\mathcal{S}_2$ {\emph{vs.}} $\alpha$ reveals that $\mathcal{S}_2\neq 0$ except at $\alpha_1 = 1.8494040837...$ (Figure \ref{fig5}). For this isolated value of $\alpha$, $\lambda_2$ has no real part at $\mathcal{O}\left(\varepsilon^2\right)$. {\bf We conjecture that small-amplitude Stokes waves of all wavenumbers and in all depths are unstable to the high-frequency instability closest to the origin}, with the possible exception of Stokes waves with $\alpha = \alpha_1$.

\begin{figure}[tb]
    \centering
    \includegraphics[height=12.7cm,width=12cm]{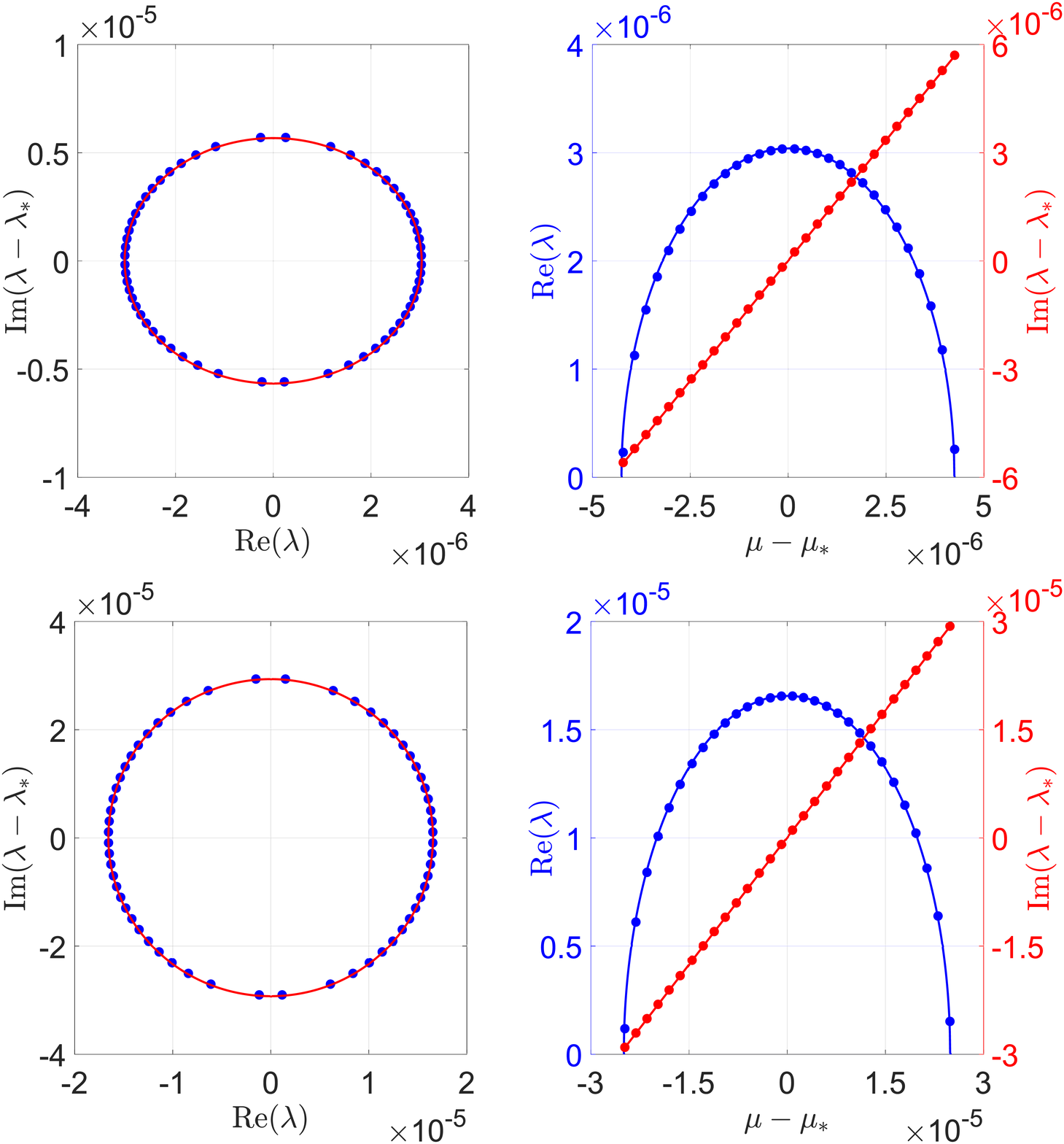}
    \caption{(Top, Left) The $p=2$ isola with $\alpha = 1.5$ and $\varepsilon = 0.01$. The most unstable eigenvalue $\lambda_{*}$ is removed from the imaginary axis for better visibility. The solid red curve is the ellipse obtained by our asymptotic calculations. The blue dots are a subset of eigenvalues from the numerically computed isola using the FFH method. (Top, Right) The Floquet parameterization of the real (blue) and imaginary (red) parts of the isola on the left. The most unstable eigenvalue $\lambda_*$ and its corresponding Floquet exponent $\mu_*$ are removed from the imaginary and Floquet axes, respectively, for better visibility. The solid curves are our asymptotic results. The colored dots are our numerical results using the FFH method. (Bottom, Left \& Right) Same with $\alpha = 1$.\\}
    \label{fig6}
\end{figure}

To $\mathcal{O}\left(\varepsilon^2\right)$, the $p=2$ isola is an ellipse in the complex spectral plane. The ellipse is constructed explicitly from the real and imaginary parts of
\begin{align}
    \lambda(\mu_2;\varepsilon) = \lambda_0 + \lambda_2(\mu_2)\varepsilon^2,
\end{align}
for $\mu_2 \in \left(M_{2,-},M_{2,+}\right)$. This ellipse has semi-major and -minor axes that are $\mathcal{O}\left(\varepsilon^2\right)$, and its center drifts from $\lambda_{0}$ along the imaginary axis like $\mathcal{O}\left(\varepsilon^2\right)$. Similarly, the interval of Floquet exponents that parameterizes this ellipse has width $\mathcal{O}\left(\varepsilon^2\right)$ and drifts from $\mu_0$ like $\mathcal{O}\left(\varepsilon^2\right)$. In Figure \ref{fig6}, we compare the ellipse with a subset of numerically computed eigenvalues on the $p=2$ isola for $\varepsilon = 0.01$ and find excellent agreement. We find similar agreement between the Floquet parameterization of the ellipse and of the numerically computed isola.    

\begin{figure}[tb]
    \centering
    \includegraphics[height=12.7cm,width=12cm]{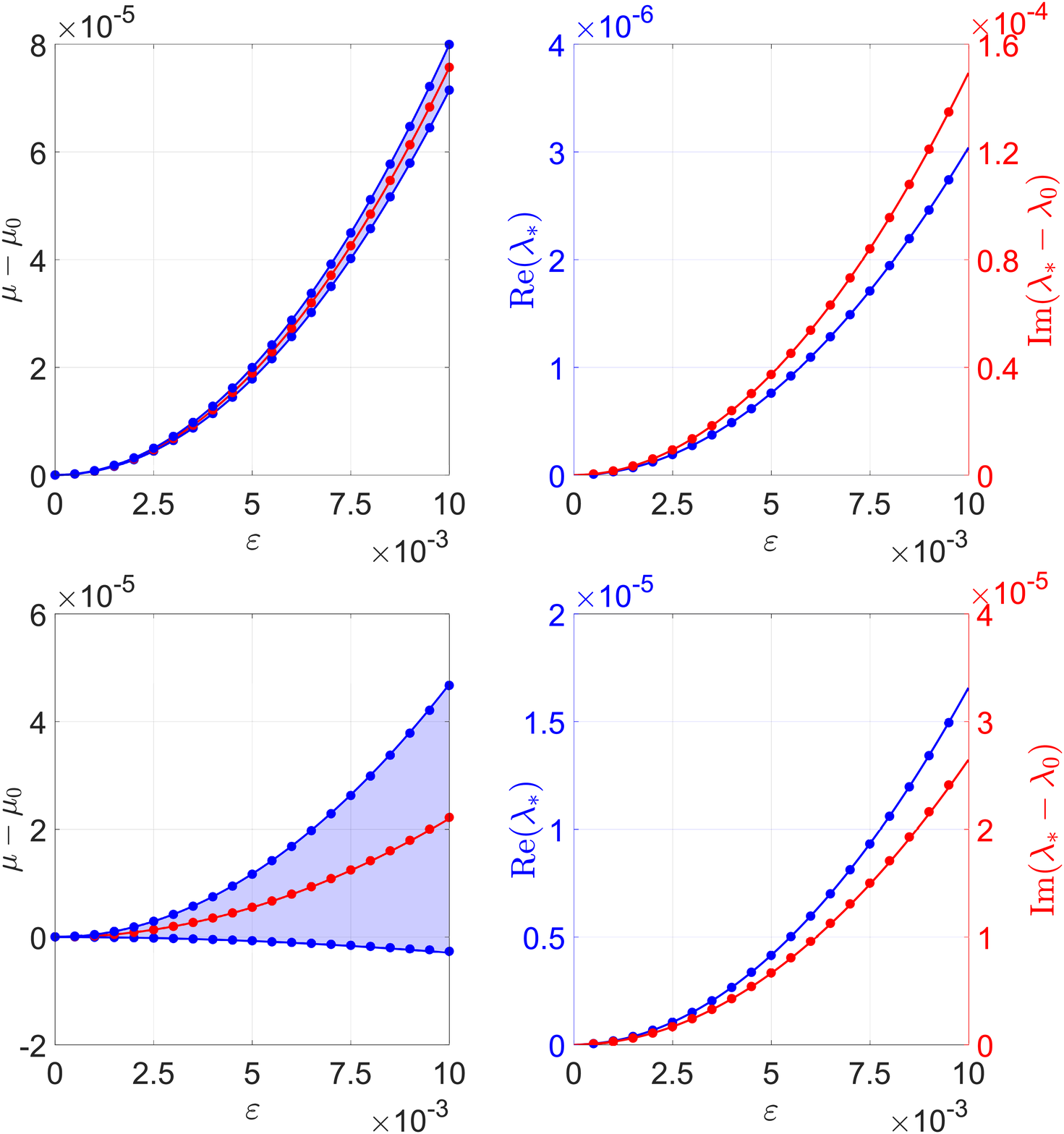}
    \caption{ (Top, Left) The interval of Floquet exponents parameterizing the $p=2$ isola as a function of $\varepsilon$ for $\alpha = 1.5$. The zeroth-order correction of the Floquet exponent is removed from the Floquet axis for better visibility. The solid blue curves are the boundaries of this interval according to our asymptotic calculations. The blue dots are the boundaries computed numerically by the FFH method. The solid red curve gives the Floquet exponent of the most unstable eigenvalue on the isola according to our asymptotic calculations. The red dots are the Floquet exponent of the most unstable eigenvalue as computed by the FFH method. (Top, Right) The real (blue) and imaginary (red) parts of the most unstable eigenvalue of the $p=2$ isola with $\alpha = 1.5$ as a function of $\varepsilon$. The zeroth-order correction of the eigenvalue is removed from the imaginary axis for better visibility. The solid curves are our asymptotic calculations. The colored dots are our numerical results using the FFH method. (Bottom, Left \& Right) Same with $\alpha = 1$.\\}
    \label{fig7}
\end{figure}

The eigenvalue of largest real part on the ellipse occurs when $\mu_2 = \mu_{2,*}$. Thus, the leading-order behavior of the most unstable eigenvalue on the $p=2$ isola has real and imaginary parts
\begin{subequations}
\begin{align}\label{lambdar2}
    &\lambda_{r,*} = \frac{\left|\mathcal{S}_2\right|}{2 \sqrt{\omega(n+\mu_0)\omega(n+p+\mu_0)}}\varepsilon^2 + \mathcal{O}\left(\varepsilon^3\right), \\
    \lambda_{i,*} = -\Omega_{1}\left(n+\mu_0\right) -& \left(\frac{\mathfrak{p}_{2,n+p}c_{g,1}\left(n+\mu_0 \right)-\mathfrak{p}_{2,n}c_{g,-1}\left(n+p+\mu_0 \right)}{c_{g,-1}\left(n+p+\mu_0\right)-c_{g,1}\left(n+\mu_0\right)}\right)\varepsilon^2 + \mathcal{O}\left(\varepsilon^3\right),
\end{align}
\end{subequations}
respectively. The corresponding Floquet exponent is
\begin{align}
    \mu_* = \mu_0 + \mu_{2,*} \varepsilon^2 +\mathcal{O}\left(\varepsilon^3\right).
\end{align}
These expansions agree well with numerical results (Figure \ref{fig7}).

\vspace*{0.1in}

\begin{figure}[tb]
    \centering
    \includegraphics[height=6.1cm,width=12cm]{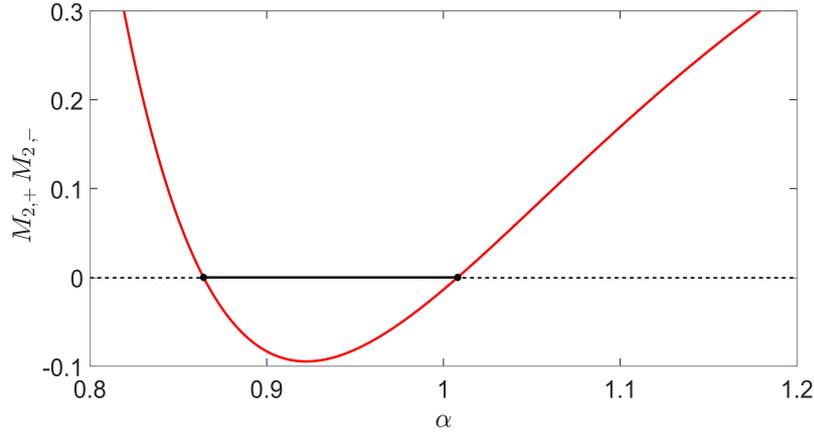}
    \caption{ A plot of $M_{2,+}M_{2,-}$ \emph{vs.} $\alpha$ (solid red). We find $M_{2,+}M_{2,-} < 0$ only when $\alpha \in \left(0.8643029367..., 1.0080416077...\right)$ (solid black). If $M_{2,+}M_{2,-} < 0$, the boundaries of the Floquet exponents parameterizing the $p=2$ isola have opposite concavities at $\varepsilon =0$. Only then does $\mu_0$ remain in the interval of Floquet exponents parameterizing the isola for positive $\varepsilon$. 
    \\}
    \label{fig7bonus}
\end{figure}

\noindent \textbf{Remark 8}. According to Figure \ref{fig7}, $\mu_0$ is contained within the interval parameterizing the $p=2$ isola if the boundaries of this interval have opposite concavity at $\varepsilon=0$. This occurs if and only if $M_{2,+}M_{2,-}<0$. In Figure \ref{fig7bonus}, we plot $M_{2,+}M_{2,-}$ as a function of $\alpha$. We find $M_{2,+}M_{2,-}<0$ only if $\alpha \in \left(0.8643029367..., 1.0080416077...\right)$. Hur \emph{\&} Yang \cite{huryang20} prove the existence of an eigenvalue with Floquet exponent $\mu_0$ on the $p=2$ isola for $\alpha$ in this interval. As we have demonstrated, to account for $p=2$ high-frequency instabilities that occur outside this interval, it is necessary to expand the Floquet exponent as a power series in $\varepsilon$ about $\mu_0$.

\subsection{The Case of Infinite Depth}
In infinite depth, the $p=2$ isola originates from the eigenvalue
\begin{align}
\lambda_0 = -\frac{3}{4}i, 
\end{align}
with corresponding Floquet exponent $\mu_0 =-1/4$ and $n = -2$, see Remark 6. The corresponding eigenfunction, after normalizing, is
\begin{align}
    {\bf w}_0 = \begin{pmatrix} 1 \\ \frac23 i \end{pmatrix}e^{inx} + \gamma_0 \begin{pmatrix} 1 \\ -2i \end{pmatrix}e^{i(n+p)x},
\end{align}
where $\gamma_0 \in \mathbb{C} \setminus \{0 \}$. We modify the generalized eigenvalue problem \eqref{24} according to Remark 5 and expand the spectral data as a power series in $\varepsilon$ about the values above.

Terms of $\mathcal{O}\left(\varepsilon^0\right)$ cancel by construction. At $\mathcal{O}\left(\varepsilon\right)$, the solvability conditions simplify to \begin{align} \lambda_1 = 0 = \mu_1, \end{align} as in finite depth, and the normalized solution of the $\mathcal{O}(\varepsilon)$ problem is
\begin{align}
    {\bf w}_1 = \sum_{\substack{j = n-1 \\ j \neq n, n+p}}^{n+p+1} \hat{\mathcal{W}}_{1,j,\infty}e^{ijx} + \gamma_1 \begin{pmatrix} 1 \\ -2i\end{pmatrix} e^{i(n+p)x},
\end{align}
where the coefficients $\hat{\mathcal{W}}_{1,j,\infty}$ depend at most linearly on $\gamma_0$. 

At $\mathcal{O}\left(\varepsilon^2\right)$, the solvability conditions are
\begin{subequations}
\begin{align}
    \lambda_2 + i\mathfrak{c}_{2,1,n,\infty} &= 0, \label{36a} \\
    \gamma_0\left(\lambda_2 + i\mathfrak{c}_{2,-1,n+p,\infty}\right) &= 0, \label{36b}
\end{align}
\end{subequations}
where 
\begin{align}\mathfrak{c}_{2,\sigma,j,\infty} = \mu_2 c_{g,\sigma,\infty}\left(j+\mu_0\right) - \mathfrak{p}_{2,j,\infty},
\end{align}
for $c_{g,\sigma,\infty}(z) = \lim_{\alpha \rightarrow \infty} \Omega_{\sigma,z}(z)$ and explicitly computed constants $\mathfrak{p}_{2,j,\infty}$.

Since $\gamma_0 \neq 0$, equations \eqref{36a}-\eqref{36b} reduce to a linear system for $\lambda_2$ and $\mu_2$. The solution of this system is
\begin{align}
    \lambda_2 = \frac{55}{32}i, \quad  \quad \mu_2 = \frac{57}{64}. 
\end{align}
Since $\lambda_2$ is purely imaginary, the leading-order behavior of the $p=2$ isola does not occur at $\mathcal{O}\left(\varepsilon^2\right)$, as expected from \eqref{lambdar2}, since $\lim_{\alpha\rightarrow \infty}\mathcal{S}_2=0$. Thus, while the asymptotic expressions involved in infinite depth are simpler than those in finite depth, {\bf the leading-order behavior of the $p=2$ isola requires a higher-order calculation in infinite depth.} We obtain the normalized solution of the $\mathcal{O}\left(\varepsilon^2\right)$ problem:
\begin{align}
    {\bf w}_2 = \sum_{\substack{j = n-2}}^{n+p+2} \hat{\mathcal{W}}_{2,j,\infty}e^{ijx} + \gamma_2 \begin{pmatrix} 1 \\ -2i\end{pmatrix} e^{i(n+p)x},
\end{align}
where the coefficients $\hat{\mathcal{W}}_{2,j,\infty}$ depend at most linearly on $\gamma_0$ and $\gamma_1$ while $\gamma_2 \in \mathbb{C}$ is a free parameter at this order. 

At $\mathcal{O}\left(\varepsilon^3\right)$, the solvability conditions reduce to 
\begin{subequations}
\begin{align}
    \lambda_3 +i
\mu_3c_{g,1,\infty}\left(n+\mu_0\right) &= 0, \\
\gamma_0\left( \lambda_3 + i
\mu_3c_{g,-1,\infty}\left(n+p+\mu_0\right)\right) &= 0.
\end{align}
\end{subequations}
As in finite depth, $c_{g,1,\infty}\left(n+\mu_0\right) \neq c_{g,-1,\infty}\left(n+p+\mu_0\right)$, and since $\gamma_0 \neq 0$, we must have
\begin{align}
    \lambda_3 = 0 = \mu_3.
\end{align}
No instability is observed at this order. The normalized solution of the $\mathcal{O}\left(\varepsilon^3\right)$ problem is
\begin{align}
      {\bf w}_3 = \sum_{\substack{j = n-3}}^{n+p+3} \hat{\mathcal{W}}_{3,j,\infty}e^{ijx} + \gamma_3 \begin{pmatrix} 1 \\ -2i\end{pmatrix} e^{i(n+p)x},
\end{align}
where the coefficients $\hat{\mathcal{W}}_{3,j,\infty}$ depend at most linearly on $\gamma_0$, $\gamma_1$, and $\gamma_2$ while the parameter $\gamma_3 \in \mathbb{C}$ is free at this order.

At $\mathcal{O}\left(\varepsilon^4\right)$, the solvability conditions are
\begin{subequations}
\begin{align}
    2\left(\lambda_4 + i\mathfrak{c}_{4,1,n,\infty}\right) + i\gamma_0\mathfrak{s}_{4,n,\infty} &= 0, \label{37a} \\
    2\gamma_0\left(\lambda_4 + i\mathfrak{c}_{4,-1,n+p,\infty}\right) + i \mathfrak{s}_{4,n+p,\infty} &= 0, \label{37b}
\end{align}
\end{subequations}
where 
\begin{align}\mathfrak{c}_{4,\sigma,j,\infty} = \mu_4 c_{g,\sigma,\infty}\left(j+\mu_0\right) - \mathfrak{p}_{4,j,\infty},
\end{align}
and $\mathfrak{s}_{4,j,\infty}$ and $\mathfrak{p}_{4,j,\infty}$ are explicitly computed constants. Substituting these constants into \eqref{37a}-\eqref{37b} and solving for $\lambda_4$, we find the explicit formula
\begin{align}
    \lambda_4 &= \frac{(48671+49152\mu_4)}{36864} i \pm \frac{\sqrt{-134933977+291053568\mu_4-150994944\mu_4^2}}{18432}. \label{38}
\end{align}
Equation \eqref{38} has nonzero real part provided
\begin{align}
    \mu_4 \in \left(\frac{11843}{12288} - \frac{111\sqrt{3}}{1024}, \frac{11843}{12288} + \frac{111\sqrt{3}}{1024} \right). \label{39}
\end{align}
Thus, the $p=2$ isola is an ellipse to $\mathcal{O}\left(\varepsilon^4\right)$ given by the real and imaginary parts of 
\begin{align}
    \lambda(\mu_4;\varepsilon) = -\frac34 i + \frac{55}{32}i\varepsilon^2 + \lambda_4(\mu_4)\varepsilon^4,
\end{align}
for $\mu_4$ in \eqref{39}. 
Unlike in finite depth, this ellipse has semi-major and -minor axes that are $\mathcal{O}\left(\varepsilon^4\right)$, while the center drifts from $\lambda_{0}$ like $\mathcal{O}\left(\varepsilon^2\right)$. Similarly, the Floquet parameterization of the isola has width $\mathcal{O}\left(\varepsilon^4\right)$ and drifts from $\mu_0$ like $\mathcal{O}\left(\varepsilon^2\right)$.

\begin{figure}[tb]
    \centering
    \includegraphics[height=6.5cm,width=14cm]{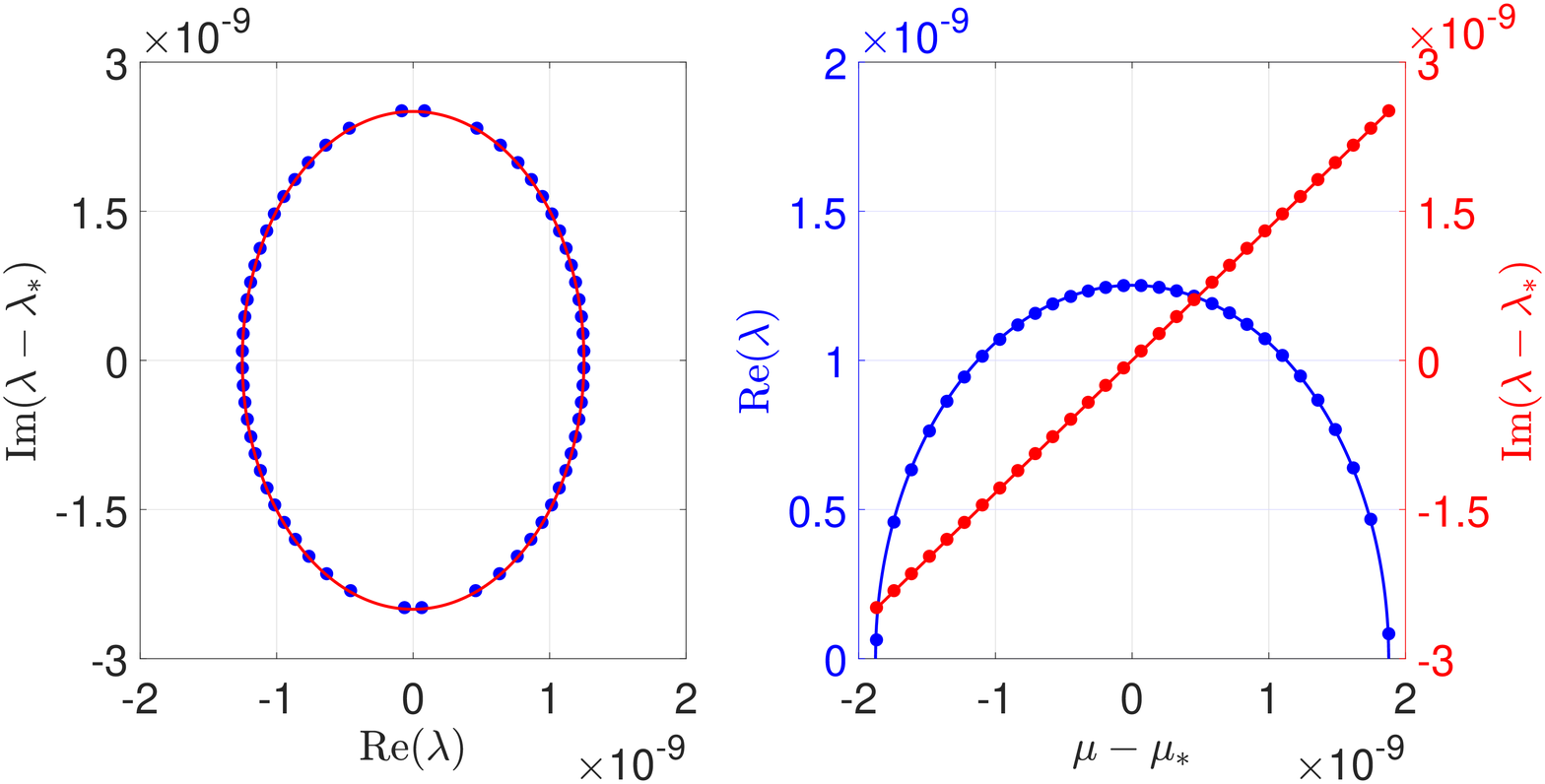}
    \caption{(Left) The $p=2$ isola with $\alpha=\infty$ and $\varepsilon = 0.01$. The most unstable eigenvalue $\lambda_{*}$ is removed from the imaginary axis for better visibility. The solid red curve is the ellipse obtained by our asymptotic calculations. The blue dots are a subset of eigenvalues from the numerically computed isola using the FFH method. (Right) The Floquet parameterization of the real (blue) and imaginary (red) parts of the isola. The most unstable eigenvalue $\lambda_*$ and its corresponding Floquet exponent $\mu_*$ are removed from the imaginary and Floquet axes, respectively, for better visibility. The solid curves are our asymptotic results. The colored dots are our numerical results using the FFH method.\\}
    \label{fig8}
\end{figure}

In Figure \ref{fig8}, we compare the asymptotically computed ellipse with a subset of numerically computed eigenvalues on the $p=2$ isola for $\varepsilon = 0.01$. Notice this ellipse is considerably smaller than that in finite depth for comparable wave amplitude (Figure \ref{fig6}). Excellent agreement is found between the asymptotic and numerical predictions. Similar agreement is found between the Floquet parameterization of the ellipse and of the numerically computed isola.

\begin{figure}[tb]
    \centering
    \includegraphics[height=6.5cm,width=14cm]{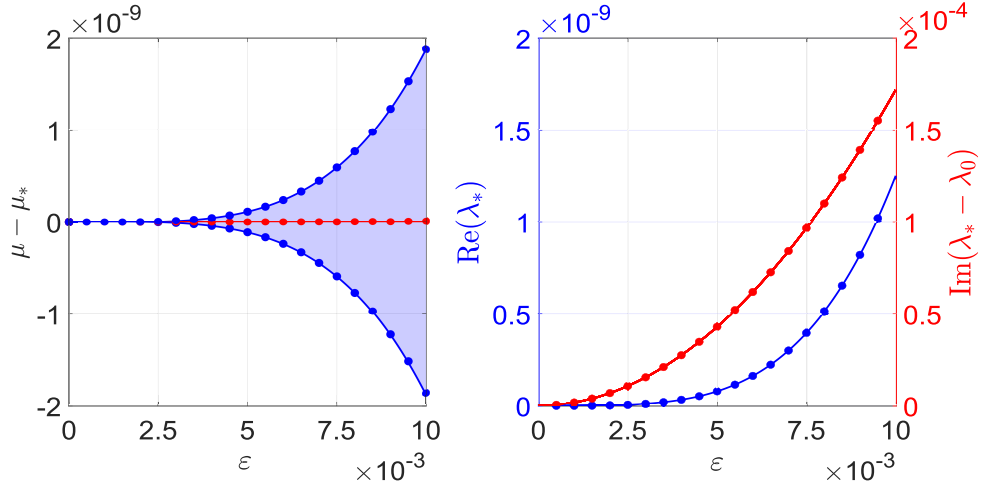}
    \caption{ (Left) The interval of Floquet exponents parameterizing the $p=2$ isola as a function of $\varepsilon$ for $\alpha = \infty$. The most unstable Floquet exponent $\mu_*$ is removed from the Floquet axis for better visibility. The solid blue curves are the boundaries of this interval according to our asymptotic calculations. The blue dots are the boundaries computed numerically by the FFH method. The solid red curve gives the Floquet exponent of the most unstable eigenvalue on the isola according to our asymptotic calculations. The red dots are the Floquet exponent of the most unstable eigenvalue as computed by the FFH method. (Right) The real (blue) and imaginary (red) parts of the most unstable eigenvalue of the $p=2$ isola with $\alpha = \infty$ as a function of $\varepsilon$. The zeroth-order correction of the eigenvalue is removed from the imaginary axis for better visibility. The solid curves are our asymptotic calculations. The colored dots are our numerical results using the FFH method. }
    \label{fig9}
\end{figure}

The eigenvalue of largest real part on the ellipse occurs when $\mu_4 = 11843/36864$. Thus, the real and imaginary parts of the most unstable eigenvalue on the isola have asymptotic expansions
\begin{align}
    \lambda_{r,*} &= \frac{37\sqrt{3}}{512} \varepsilon^4 + \mathcal{O}\left(\varepsilon^5\right), \\
    \lambda_{i,*} &= -\frac34 + \frac{55}{32} \varepsilon^2 + \frac{96043}{36864} \varepsilon^4 + \mathcal{O}\left(\varepsilon^5\right),
\end{align}
respectively. The corresponding Floquet exponent has expansion
\begin{align}
    \mu_* = -\frac14 + \frac{57}{64}\varepsilon^2 + \frac{11843}{36864}\varepsilon^4 + \mathcal{O}\left(\varepsilon^5\right).
\end{align}
These expansions are compared with numerical results in Figure \ref{fig9}.

\section{Second Isola. High-Frequency Instabilities: $p=3$}

We extend the perturbation method developed in Section 5 to obtain the leading-order behavior of the high-frequency isola that arises from $\lambda_{0,p}$ with $p=3$. This isola is the second closest to the origin by Theorem 1, and its leading-order behavior is obtained at $\mathcal{O}\left(\varepsilon^3\right)$. 

As in the previous section, we expand the spectral data of \eqref{24} according to \eqref{29a}-\eqref{29c} and normalize the eigenfunctions according to \eqref{28.75} for convenience. The perturbation method proceeds as in Section 5, with two major changes:
\begin{enumerate}
    \item[(i)] At $\mathcal{O}\left(\varepsilon^2\right)$, the solvability conditions are independent of $\gamma_0$ and linear in $\lambda_2$ and $\mu_2$. As a consequence, $\lambda_2$ is purely imaginary, and the leading-order behavior of the isola is undetermined at this order.
    \item[(ii)] At $\mathcal{O}\left(\varepsilon^3\right)$, the solvability conditions depend on $\gamma_0$, $\lambda_3$, and $\gamma_1$. Using solvability conditions from the previous order together with the collision condition \eqref{27}, one shows that the dependence on $\gamma_1$ vanishes from these conditions.
\end{enumerate}
A more complete description of these calculations is provided in Appendix B.

\subsection{The $\mathcal{O}\left(\varepsilon^3\right)$ Problem}

\begin{figure}[tb]
     \hspace*{-0.0cm}
    \includegraphics[height=7.5cm,width=15.5cm]{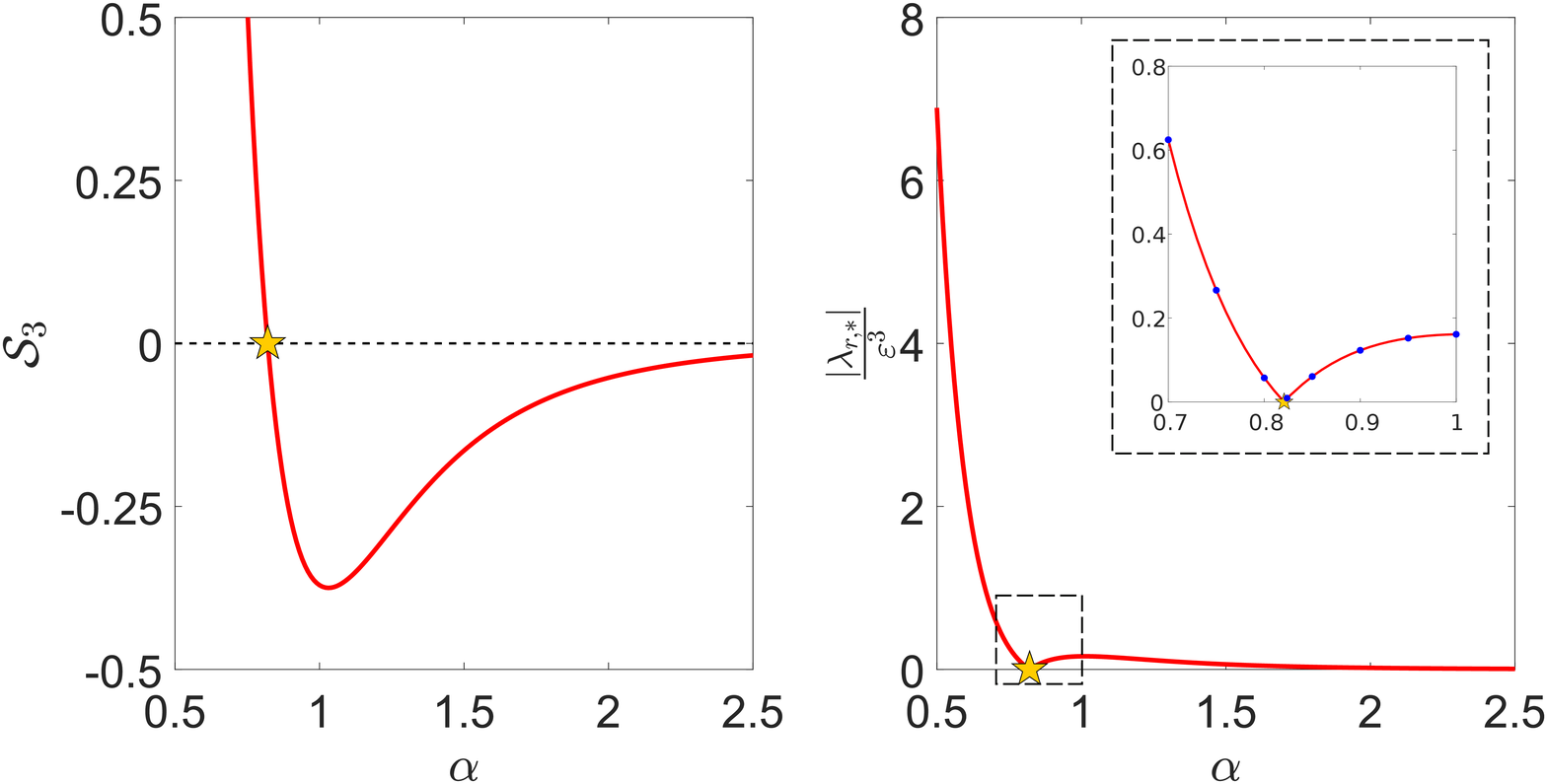} \vspace*{0.15cm}
    \caption{(Left) A plot of $\mathcal{S}_3$ \emph{vs.} $\alpha$ (solid red). The zero of $\mathcal{S}_3$ for $\alpha > 0$ is $\alpha_2 = 0.8206431673...$ (gold star). (Right) The real part $\lambda_{r,*}$ of the most unstable eigenvalue on the $p=3$ isola as a function of $\alpha$ according to our asymptotic calculations (solid red). The real part of the eigenvalue is normalized by $\varepsilon^3$ for better visibility. We zoom-in around $\alpha = \alpha_2$ (gold star) in the inlay. The real part of the most unstable eigenvalue on the isola vanishes as $\alpha \rightarrow \alpha_2$ according to our asymptotic calculations, which agrees with our numerical results using the FFH method with $\varepsilon = 0.01$ (blue dots).\\}
    \label{fig10}
\end{figure}

Solving for $\lambda_3$ in the solvability conditions at $\mathcal{O}\left(\varepsilon^3\right)$, we find
\begin{align}
   \lambda_3 &= -i\mu_3\biggr( \frac{c_{g,-1}(n+p+\mu_0) + c_{g,1}(n+\mu_0)}{2} \biggr)  \label{42} \\ &\quad \quad \quad \pm \sqrt{-\mu_3^2\biggr(\frac{c_{g,-1}(n+p+\mu_0) -c_{g,1}(n+\mu_0)}{2} \biggr)^2 + \frac{\mathcal{S}_3^2}{4\omega(n+\mu_0)\omega(n+p+\mu_0)} }. \nonumber
\end{align}
Similar to $\mathcal{S}_2$ in the previous section, $\mathcal{S}_3$ is another lengthy expression depending only on $\alpha$, see Appendix B for more details. A plot of $\mathcal{S}_3$ \emph{vs.} $\alpha$ reveals $\mathcal{S}_3 \neq 0$, except at $\alpha_2 = 0.8206431673...$ (Figure \ref{fig10}). {\bf We conjecture that Stokes waves of all wavenumbers and in all depths are unstable to the second closest high-frequency instability from the origin}, with possible exceptions if $\alpha = \alpha_2$. Since $\alpha_2\neq \alpha_1$, Stokes waves of all wavenumbers and in all depths appear to be unstable with respect to high-frequency instabilities.

\vspace*{0.1in}

\noindent \textbf{Remark 9}. As $\alpha \rightarrow \infty$, $\mathcal{S}_3 \rightarrow 0$. Therefore, the leading-order behavior of the $p=3$ isola in infinite depth is resolved at higher order, similar to the $p=2$ case. For $\varepsilon$ on the order of 0.01, this isola is already within the numerical error of the FFH method. For larger $\varepsilon$, the expansions deviate too quickly from the numerics to make comparisons.  

\vspace*{0.1in}

Provided $\alpha \neq \alpha_2$, \eqref{42} has nonzero real part for $\mu_3 \in \left(-M_3,M_3\right)$, where
\begin{align}
    M_3 &= \frac{|\mathcal{S}_3|}{|c_{g,-1}(n+p+\mu_0) - c_{g,1}(n+\mu_0)|\sqrt{\omega(n+\mu_0)\omega(n+p+\mu_0)}}. \label{52}
\end{align}
Unlike the $p=2$ isola, this interval is symmetric about the origin. For $\mu_3$ in this interval, the real and imaginary parts of \eqref{42}, together with the lower-order corrections of $\lambda$, trace an ellipse asymptotic to the $p=3$ isola. This ellipse has semi-major and -minor axes that scale as $\mathcal{O}\left(\varepsilon^3\right)$ and a center that drifts form $\lambda_0$ like $\mathcal{O}\left(\varepsilon^2\right)$. The Floquet parameterization of this ellipse has width $\mathcal{O}\left(\varepsilon^3\right)$ and drifts from $\mu_0$ like $\mathcal{O}\left(\varepsilon^2\right)$. As a result, this isola is more challenging to capture than the $p=2$ isola in finite depth. 

\begin{figure}[tb]
    \centering
    \includegraphics[height=12.7cm,width=12cm]{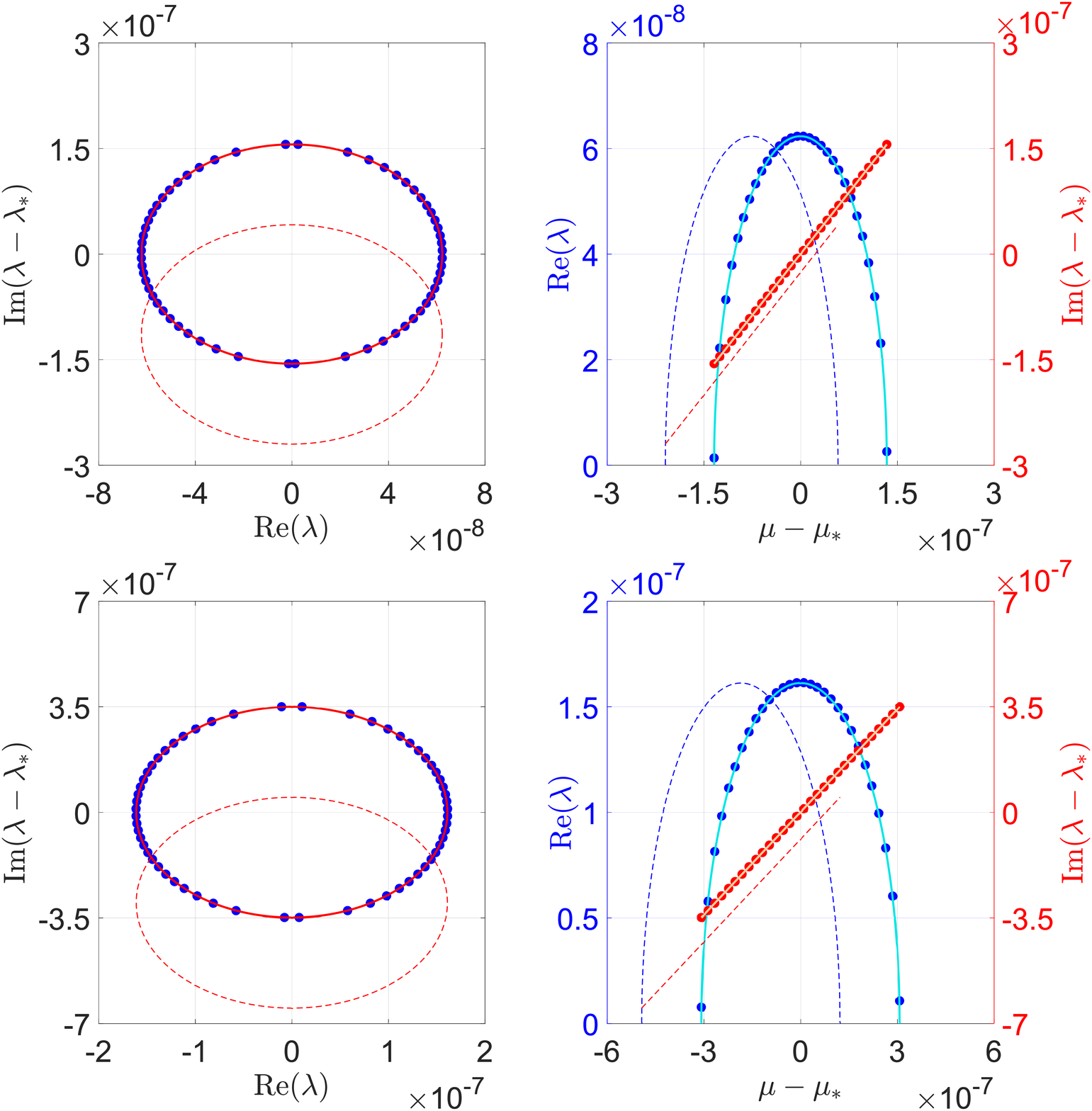}
    \caption{(Top, Left) The $p=3$ isola with $\alpha = 1.5$ and $\varepsilon = 0.01$. The most unstable eigenvalue $\lambda_{*}$ is removed from the imaginary axis for better visibility. The solid and dashed red curves are the ellipses obtained by our $\mathcal{O}\left(\varepsilon^4\right)$ and $\mathcal{O}\left(\varepsilon^3\right)$ asymptotic calculations, respectively. The blue dots are a subset of eigenvalues from the numerically computed isola using the FFH method. (Top, Right) The Floquet parameterization of the real (blue) and imaginary (red) parts of the isola on the left. The most unstable eigenvalue $\lambda_*$ and its corresponding Floquet exponent $\mu_*$ are removed from the imaginary and Floquet axes, respectively, for better visibility. The solid teal and orange curves are our asymptotic results for the real and imaginary parts of the Floquet parameterization, respectively, to $\mathcal{O}\left(\varepsilon^4\right)$. The dashed blue and red curves are the same results to $\mathcal{O}\left(\varepsilon^3\right)$. The blue and red dots are the numerically computed real and imaginary parts of the Floquet parameterization, respectively, using the FFH method. (Bottom, Left \& Right) Same with $\alpha = 1$.\\}
    \label{fig11}
\end{figure}

Comparing our asymptotic and numerical $p=3$ isolas with $\varepsilon = 0.01$ (Figure \ref{fig11}), we observe that, while the real part of the numerical isola matches our $\mathcal{O}\left(\varepsilon^3\right)$ calculations, the imaginary part and Floquet parameterization of the isola require fourth-order corrections. This is in contrast with the $p=2$ isola (Figure \ref{fig6}), for which we obtain the drifts in the imaginary part and Floquet parameterization at the same order as the real part. We obtain these drifts for the $p=3$ isola in the following subsection. 

Equating $\mu_3 = 0$ maximizes the real part of \eqref{42}. Hence, the real and imaginary part of the most unstable eigenvalue on the $p=3$ isola have asymptotic expansions 
\begin{subequations}
\begin{align}
    \lambda_{r,*} &=  \left( \frac{|\mathcal{S}_3|}{2\sqrt{\omega(n+\mu_0)\omega(n+p+\mu_0)}}\right)\varepsilon^3 + \mathcal{O}\left(\varepsilon^4\right), \label{48a} \\
    \lambda_{i,*} &= -i\Omega_1(n+\mu_0) -\left(\frac{\mathfrak{p}_{2,n+p}c_{g,1}(n+\mu_0) - \mathfrak{p}_{2,n}c_{g,-1}(n+p+\mu_0)}{c_{g,-1}(n+p+\mu_0)-c_{g,1}(n+\mu_0)}\right)\varepsilon^2  +\mathcal{O}\left(\varepsilon^4\right), \label{48b} 
\end{align}
\end{subequations}
respectively, and the corresponding Floquet exponent has asymptotic expansion
\begin{align}
    \mu_* = \mu_0 + \left(\frac{\mathfrak{p}_{2,n+p}-\mathfrak{p}_{2,n}}{c_{g,-1}(n+p+\mu_0)-c_{g,1}(n+\mu_0)}\right)\varepsilon^2 + \mathcal{O}\left(\varepsilon^4\right). \label{48c}
\end{align}
The quantities $\mathfrak{p}_{2,j}$ are defined in Appendix B.

Figure \ref{fig12} compares the asymptotic expansions \eqref{48a}-\eqref{48b} and \eqref{48c} with their numerical counterparts. Excellent agreement is found for the real and imaginary parts of the most unstable eigenvalue. The interval of Floquet exponents that parameterizes the isola requires a fourth-order correction to match the numerical predictions. 
\begin{figure}[tb!]
    \centering
    \includegraphics[height=12.7cm,width=12cm]{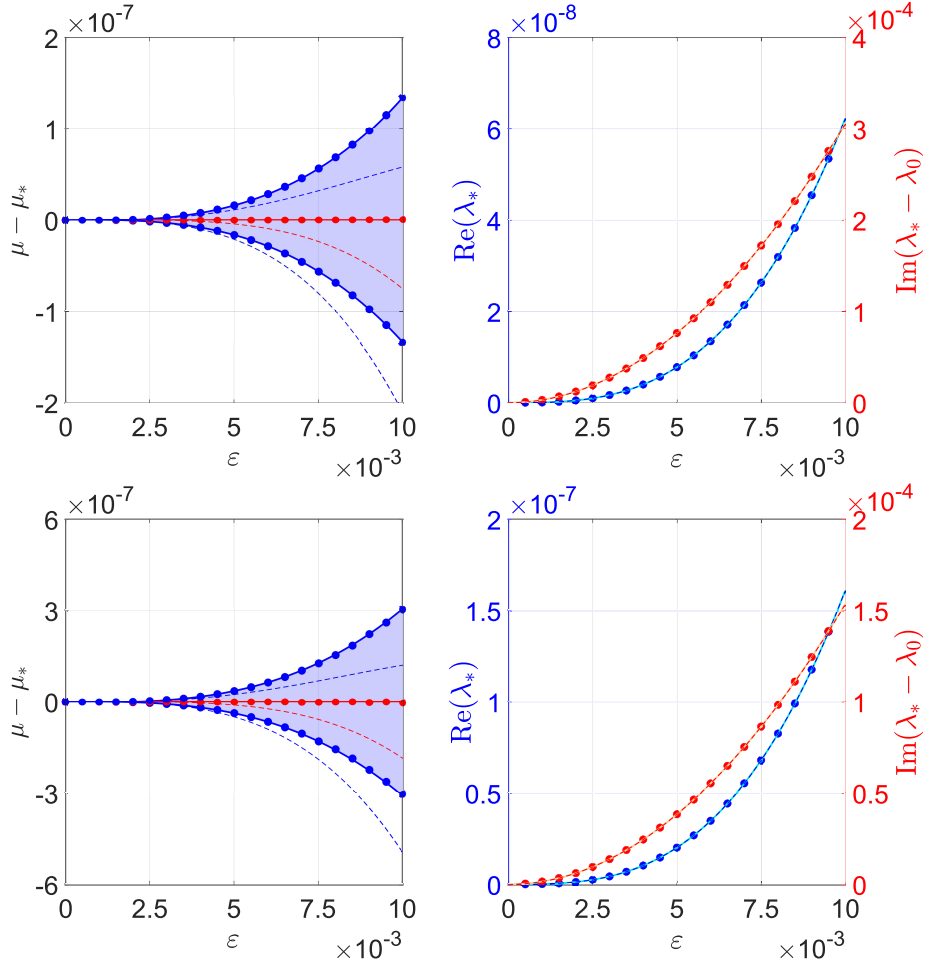}
    \caption{(Top, Left) The interval of Floquet exponents parameterizing the $p=3$ isola as a function of $\varepsilon$ for $\alpha = 1.5$. The most unstable Floquet exponent $\mu_*$ is removed from the Floquet axis for better visibility. The solid and dashed blue curves are the boundaries of this interval according to our $\mathcal{O}\left(\varepsilon^4\right)$ and  $\mathcal{O}\left(\varepsilon^3\right)$ asymptotic calculations, respectively. The blue dots are the boundaries computed numerically by the FFH method. The solid and dashed red curves give the Floquet exponent of the most unstable eigenvalue on the isola according to our  $\mathcal{O}\left(\varepsilon^4\right)$ and  $\mathcal{O}\left(\varepsilon^3\right)$ asymptotic calculations, respectively. The red dots are the Floquet exponent of the most unstable eigenvalue as computed by the FFH method. (Top, Right) The real (blue) and imaginary (red) parts of the most unstable eigenvalue of the $p=3$ isola with $\alpha = 1.5$ as a function of $\varepsilon$. The zeroth-order correction of the eigenvalue is removed from the imaginary axis for better visibility. The solid teal and orange curves are our asymptotic calculations for the real and imaginary parts of the most unstable eigenvalue to $\mathcal{O}\left(\varepsilon^4\right)$, respectively. The dashed blue and red curves are the same results to $\mathcal{O}\left(\varepsilon^3\right)$. The blue and red dots are the numerically computed real and imaginary parts of the most unstable eigenvalue using the FFH method. (Bottom, Left \& Right) Same with $\alpha = 1$.\\ }
    \label{fig12}
\end{figure}

Before proceeding to the next order, we solve the $\mathcal{O}\left(\varepsilon^3\right)$ problem for ${\bf w}_3$:
 \begin{align}
    {\bf w}_3 = \sum_{\substack{j = n-3} }^{n+p+3} \hat{\mathcal{W}}_{3,j} e^{ijx} + \gamma_3 \begin{pmatrix} 1 \\ \frac{i}{\omega(n+p+\mu_0)} \end{pmatrix} e^{i(n+p)x},
\end{align}
where the coefficients $\hat{\mathcal{W}}_{3,j}$ depend on $\alpha$ (possibly through intermediate dependencies on known zeroth-, first-, and second-order results) and at most linearly on $\gamma_0, \gamma_1,$ and $\gamma_2$. At this order, $\gamma_3 \in \mathbb{C}$ is a free parameter.
\subsection{The $\mathcal{O}\left(\varepsilon^4\right)$ Problem}
At $\mathcal{O}\left(\varepsilon^4\right)$, the spectral problem \eqref{24} becomes
\begin{align} {
    \left(L_0 - \lambda_0R_0 \right){\bf w}_4 = \left(\sum_{j=0}^2\lambda_{4-j} R_j\right){\bf w}_0 + \left( \sum_{j=0}^1 \lambda_{3-j} R_j\ \right){\bf w}_1 + \lambda_2R_0{\bf w}_2 - \sum_{j=0}^{3}\left(L_{4-j}-\lambda_0R_{4-j} \right){\bf w}_j.} \label{50}
\end{align}
After some manipulation, the solvability conditions of \eqref{50} can be written as
\begin{align} {\small
    \begin{pmatrix} 2 & i\mathfrak{s}_{3,n} \\ 2\gamma_0 & 2\left(\lambda_3 +i\mu_3c_{g,-1}(n+p+\mu_0) \right) \end{pmatrix} \begin{pmatrix} \lambda_4 \\ \gamma_1 \end{pmatrix} + i\gamma_2 \begin{pmatrix} 0 \\ \mathfrak{t}_{4,n+p}\end{pmatrix} = -2i\begin{pmatrix} \mu_4c_{g,1}(n+\mu_0)-\mathfrak{p}_{4,n} \\ \gamma_0\left(\mu_4c_{g,-1}(n+p+\mu_0) - \mathfrak{p}_{4,n+p} \right) \end{pmatrix}.} \label{51}
\end{align}
Using the solvability conditions at the previous order and the collision condition \eqref{27}, one can show $\mathfrak{t}_{4,n+p} \equiv 0$. Then, \eqref{51} reduces to a linear system for $\lambda_4$ and $\gamma_1$.

For $\mu_3 \in \left(-M_3,M_3\right)$ with $M_3$ given by \eqref{52}, the determinant of \eqref{51} simplifies to
\begin{align}
    \textrm{det}\begin{pmatrix} 2 & i\mathfrak{s}_{3,n} \\ 2\gamma_0 & 2\left(\lambda_3 +i\mu_3c_{g,-1}(n+p+\mu_0) \right) \end{pmatrix} = 8\lambda_{3,r},
\end{align}
where $\lambda_{3,r} = \textrm{Re}\left(\lambda_3\right).$ Provided $\alpha \neq \alpha_2$, \eqref{51} is invertible for all $\mu_3 \in \left(-M_3,M_3\right)$.

Solving \eqref{51} for $\lambda_4$, 
\begin{align}
    \lambda_4 =& i\biggr[\frac{\left(\lambda_3 + i\mu_3c_{g,-1}(n+p+\mu_0) \right)\left(c_{g,1}(n+\mu_0)-\mathfrak{p}_{4,n} \right)}{2\lambda_{3,r}} \phantom] \\ 
    &\quad \quad \phantom[ + \frac{\left(\lambda_3 + i\mu_3c_{g,1}(n+\mu_0) \right)\left(c_{g,-1}(n+p+\mu_0)-\mathfrak{p}_{4,n+p} \right)}{2\lambda_{3,r}} \biggr]. \nonumber
\end{align}
Since $\mathfrak{p}_{4,j}, \mu_4 \in \mathbb{R}$, the real and imaginary parts of $\lambda_4=\lambda_{4,r}+i \lambda_{4,i}$ are
\begin{subequations}
\begin{align}
    \lambda_{4,r} &=  \frac{\mu_3\left(c_{g,-1}(n+p+\mu_0)-c_{g,1}(n+\mu_0) \right)}{4\lambda_{3,r}}\Big[ -\mu_4\left(c_{g,-1}(n+p+\mu_0) - c_{g,1}(n+\mu_0)\right)\phantom] \label{51a} \\ &\hspace*{7cm}+ \phantom[ \mathfrak{p}_{2,n+p} - \mathfrak{p}_{2,n} \Big], \nonumber \\
    \lambda_{4,i} &= -\frac12\Big[\mu_4\left(c_{g,-1}(n+\mu_0) + c_{g,1}(n+\mu_0)\right) - \left(\mathfrak{p}_{4,n+p} + \mathfrak{p}_{4,n} \right) \Big]. \label{51b}
\end{align}
\end{subequations}

Given \eqref{51a}-\eqref{51b}, we invoke the \emph{regular curve condition}, first introduced in Creedon \emph{ et al.} (2021a) and Creedon \emph{ et al.} (2021b). According to this condition, all eigenvalue corrections must be bounded over the closure of $\mu_3 \in \left(-M_3,M_3\right)$. Notice $\lambda_{3,r} \rightarrow 0$ as $|\mu_3| \rightarrow M_3 $. Thus, $\lambda_{4,r}$ is bounded only if 
\begin{align}
    \mu_4 = \frac{\mathfrak{p}_{4,n+p}-\mathfrak{p}_{4,n}}{c_{g,-1}(n+p+\mu_0)-c_{g,1}(n+\mu_0)}. \label{53}
\end{align}
Hence,
\begin{align}
    \lambda_4 &= -i\left(\frac{\mathfrak{p}_{4,n+p}c_{g,1}(n+\mu_0) - \mathfrak{p}_{4,n}c_{g,-1}(n+p+\mu_0)}{c_{g,-1}(n+p+\mu_0)-c_{g,1}(n+\mu_0)}\right). \label{54}
\end{align} 

\noindent \textbf{Remark 10}. If $\alpha = \alpha_2$, then $\mathfrak{s}_{3,n} =0$ and $\lambda_3 = 0 = \mu_3$. Applying the Fredholm alternative to \eqref{51}, one arrives at \eqref{53} and \eqref{54}, but the constants $\gamma_0$ and $\gamma_1$ remain arbitrary at this order. 

\vspace*{0.1in}

Equations \eqref{53} and \eqref{54} give the fourth-order drifts in the Floquet parameterization and imaginary part of the $p=3$ isola, respectively. The eigenvalues asymptotic to this isola form the ellipse
\begin{align}
    \lambda(\mu_3;\varepsilon) = \lambda_0 + \lambda_2\varepsilon^2 + \lambda_3(\mu_3)\varepsilon^3 + \lambda_4 \varepsilon^4,
\end{align}
which agrees better with the numerically computed isola than at the previous order, see Figures \ref{fig11} and \ref{fig12}.

\section{Conclusion}
Building on previous work by Akers \cite{akers15} and Creedon \emph{et al.} \cite{creedonetal21a, creedonetal21b}, we have developed a formal perturbation method to compute high-frequency instabilities of small-amplitude Stokes wave solutions of Euler's equations in arbitrary depth. This method allows one to approximate an entire high-frequency isola, going beyond standard eigenvalue perturbation methods \cite{kato66}.

We explicitly obtain the leading-order behavior of the isolas closest to the origin in the complex spectral plane ($p=2,3$) for all depths, including
\begin{enumerate}
    \item[(i)] the Floquet exponents that parameterize the isola,
    \item[(ii)] the real and imaginary parts of the most unstable eigenvalue on the isola, and
    \item[(iii)] the curve asymptotic to the isola.
\end{enumerate}
These expressions are compared directly with numerical computations of the isolas using the FFH method. Excellent agreement is found for the $p=2$ isola. The $p=3$ isola achieves similar agreement if higher-order corrections of the imaginary part and Floquet parameterization are computed using the regular curve condition, as defined in Section 6.  

According to our asymptotic results, Stokes waves of all aspect ratios, except $\kappa h = \alpha_1$ and $\kappa h = \alpha_2$, are unstable to the $p=2$ and $p=3$ high-frequency instabilities, respectively. Stokes waves are also unstable to high-frequency instabilities in infinite depth ($h=\infty$), although this requires a higher-order calculation than in finite depth. Based on these findings, we conjecture that {\bf Stokes waves of all depths and all wavenumbers are spectrally unstable to high-frequency instabilities}, extending recent work by Hur and Yang \cite{huryang20}, where the existence of the $p=2$ high-frequency instability is proven only if $\kappa h \in \left(0.86430...,1.00804...\right)$. The effect of the high-frequency instabilities on the Stokes waves has been illustrated in \cite{deconinckoliveras11}.

The perturbation method developed in this work is readily extended to higher-order isolas $(p \geq 4)$. It appears this method yields the first real-part correction of the isola at $\mathcal{O}\left(\varepsilon^p\right)$. In contrast, corrections to the imaginary part and Floquet parameterization of the isola appear at $\mathcal{O}\left(\varepsilon^2\right)$. Thus, we expect isolas further from the origin to have increasingly smaller widths, while their centers drift along the imaginary axis like $\mathcal{O}\left(\varepsilon^2\right)$. 

If correct, this conjecture highlights one of the primary challenges for analytical and numerical investigations of high-frequency instabilities: each isola is smaller than the previous, and each isola drifts from its known zeroth-order behavior quickly relative to its size. Our hope is that the perturbation method developed in this work can be used as a starting point for future proofs of high-frequency instabilities as well as improvements to the numerical resolution of high-frequency isolas far away from the origin in the complex spectral plane. 

\appendix

\section{Stokes Wave Expansions}

The Stokes waves of \eqref{7a}-\eqref{7b} have velocity 
\begin{align}
    c(\varepsilon) &= c_0 + c_2\varepsilon^2 + c_4\varepsilon^4 + \mathcal{O}\left(\varepsilon^6\right),
\end{align}
where 
\begin{subequations}
\begin{align}
    c_0^2 &= \tanh(\alpha), \\
    c_2 &= \frac{6+2\cosh(2\alpha)+\cosh(4\alpha)}{16c_0\sinh^3(\alpha)\cosh(\alpha)},\\
    c_4 &= \frac{212 + 55\cosh(2\alpha)-98\cosh(4\alpha)-23\cosh(6\alpha)+14\cosh(8\alpha)+2\cosh(10\alpha)}{2048c_0\sinh^9(\alpha)\cosh(\alpha)},
\end{align}
\end{subequations}
and take the form
\begin{subequations}
\begin{align}
 \eta_S(x;\varepsilon) &= \varepsilon \cos(x) + \varepsilon^2\hat{N}_{2,2}\cos(2x) + \varepsilon^3\hat{N}_{3,3}\cos(3x) + \varepsilon^4\Big(\hat{N}_{4,2}\cos(2x) + \hat{N}_{4,4}\cos(4x) \Big) \\ &\quad \quad + \mathcal{O}\left(\varepsilon^5\right), \nonumber \\
 q_{S,x}(x;\varepsilon) &= \frac{\varepsilon}{c_0}\cos(x) + \varepsilon^2\Big(\hat{Q}_{2,0} + \hat{Q}_{2,2}\cos(2x) \Big) + \varepsilon^3 \Big(\hat{Q}_{3,0}\cos(x) + \hat{Q}_{3,3}\cos(3x)\Big) \\ &\quad \quad + \varepsilon^4\Big(\hat{Q}_{4,0} + \hat{Q}_{4,2}\cos(2x) + \hat{Q}_{4,4}\cos(4x) \Big) + \mathcal{O}\left(\varepsilon^5\right), \nonumber
\end{align}
\end{subequations}
where 
\begin{subequations}
\allowdisplaybreaks
\begin{flalign}
\hat{N}_{2,2} &= \frac{5\cosh(\alpha)+\cosh(3\alpha)}{8\sinh^3(\alpha)}, \\
\hat{N}_{3,3} &= \frac{3(14+15\cosh(2\alpha)+6\cosh(4\alpha)+\cosh(6\alpha))}{256\sinh^6(\alpha)}, \\
\hat{N}_{4,2} &= \frac{215-418\cosh(2\alpha)-472\cosh(4\alpha)+10\cosh(6\alpha)+17\cosh(8\alpha)}{3072c_0^2\sinh^8(\alpha)}, \\
\hat{N}_{4,4} &= \frac{203+347\cosh(2\alpha)+158\cosh(4\alpha) + 76\cosh(6\alpha)+23\cosh(8\alpha)+3\cosh(10\alpha)}{768c_0^2(2+3\cosh(2\alpha))\sinh^8(\alpha)}, \\
\hat{Q}_{2,0} &= \frac{1}{4\sinh^2(\alpha)}, \\
\hat{Q}_{2,2} &= \frac{3+2\cosh(2\alpha)+\cosh(4\alpha)}{8c_0\sinh^3(\alpha)\cosh(\alpha)}, \\
\hat{Q}_{3,1} &= -\left(\frac{\cosh(2\alpha)\left(2 + \cosh(2\alpha)\right)}{16c_0\sinh^4(\alpha)} \right), \\
\hat{Q}_{3,3} &= \frac{3\left(26 - 3\cosh(2\alpha)+10\cosh(4\alpha)+3\cosh(6\alpha) \right)}{256c_0\sinh^6(\alpha)}, \\
 \hat{Q}_{4,0} &=  \frac{48+47\cosh(2\alpha)-20\cosh(4\alpha)-3\cosh(6\alpha)}{512c_0\sinh^7(\alpha)\cosh(\alpha)}, \\
\hat{Q}_{4,2} &= -\left(\frac{240+82\cosh(2\alpha)+688\cosh(4\alpha)+309\cosh(6\alpha) - 16\cosh(8\alpha) - 7\cosh(10\alpha)}{6144c_0\sinh^9(\alpha)\cosh(\alpha)} \right),\\
    \hat{Q}_{4,4} &= \frac{408+638\cosh(2\alpha)+230\cosh(4\alpha) + 171\cosh(6\alpha) + 124\cosh(8\alpha)+43\cosh(10\alpha) + 6\cosh(12\alpha)}{1536c_0(2+3\cosh(2\alpha))\sinh^9(\alpha)\cosh(\alpha)}.
\end{flalign}
\end{subequations}
The Stokes expansions in infinite depth are obtained from the above with $\alpha \rightarrow \infty$.

\section{Detailed Calculations of the $p=3$ Instability}

For explicit representations of the asymptotic expressions derived in this appendix, see the Data Availability Statement at the end of this manuscript. 

\subsection{The $\mathcal{O}\left(\varepsilon\right)$ Problem}

At $\mathcal{O}\left(\varepsilon\right)$, the spectral problem takes the form \eqref{30}. The solvability conditions simplify to
\begin{align}
    \lambda_1 = 0 = \mu_1, \label{99}
\end{align}
and the normalized solution of the $\mathcal{O}\left(\varepsilon\right)$ problem is
\begin{align}
    {\bf w}_1 = \sum_{\substack{j = n-1 \\ j \neq n, n+p}}^{n+p+1} \hat{\mathcal{W}}_{1,j} e^{ijx} + \gamma_1 \begin{pmatrix} 1 \\ \frac{i}{\omega(n+p+\mu_0)} \end{pmatrix} e^{i(n+p)x},
    \end{align}
    where the coefficients $\hat{\mathcal{W}}_{1,j}$ depend on $\alpha$ (possibly through intermediate dependencies on known zeroth-order results) and at most linearly on $\gamma_0$. At this order, $\gamma_1 \in \mathbb{C}$ is a free parameter.
    \subsection{The $\mathcal{O}\left(\varepsilon^2\right)$ Problem}
    At $\mathcal{O}\left(\varepsilon^2\right)$, the spectral problem takes the form \eqref{32}. The solvability conditions are
    \begin{subequations}
\begin{align}
    \lambda_2 + i\mathfrak{c}_{2,1,n} &= 0, \label{100a} \\
    \gamma_0\left(\lambda_2 + i\mathfrak{c}_{2,-1,n+p}\right) &= 0, \label{100b}
\end{align}
\end{subequations}
where $\mathfrak{c}_{2,\sigma,j} = \mu_2 c_{g,\sigma}\left(j+\mu_0\right) - \mathfrak{p}_{2,j}$, as in Section 5 (although the quantities $\mathfrak{p}_{2,j}$ evaluate differently than those for the $p=2$ isolas). Since $\gamma_0 \neq 0$, the solution of \eqref{100a}-\eqref{100b} is
\begin{subequations}
\begin{align}
    \lambda_2 &= -i\left(\frac{\mathfrak{p}_{2,n+p}c_{g,1}(n+\mu_0) - \mathfrak{p}_{2,n}c_{g,-1}(n+p+\mu_0)}{c_{g,-1}(n+p+\mu_0)-c_{g,1}(n+\mu_0)}\right),\\
    \mu_2 &= \frac{\mathfrak{p}_{2,n+p}-\mathfrak{p}_{2,n}}{c_{g,-1}(n+p+\mu_0)-c_{g,1}(n+\mu_0)}.
    \end{align}
\end{subequations}
Since $\lambda_2$ is purely imaginary, no instabilities are found at this order. The normalized solution of the $\mathcal{O}\left(\varepsilon^2\right)$ problem is
\begin{align}
    {\bf w}_2 = \sum_{\substack{j = n-2} }^{n+p+2} \hat{\mathcal{W}}_{2,j} e^{ijx} + \gamma_2 \begin{pmatrix} 1 \\ \frac{i}{\omega(n+p+\mu_0)} \end{pmatrix} e^{i(n+p)x},
\end{align}
where the coefficients $\hat{\mathcal{W}}_{2,j}$ depend on $\alpha$ (possibly through intermediate dependencies on known zeroth- and first-order results) and at most linearly on $\gamma_0$ and $\gamma_1$. At this order, $\gamma_2 \in \mathbb{C}$ is a free parameter.
\subsection{The $\mathcal{O}\left(\varepsilon^3\right)$ Problem}
At $\mathcal{O}\left(\varepsilon^3\right)$, the spectral problem becomes
\begin{align}
    \left(L_0-\lambda_0R_0 \right){\bf w}_3 = \left(\lambda_2 R_1 + \lambda_3 R_0\right){\bf w}_0 - \sum_{j=0}^{2}\left(L_{3-j} -\lambda_0R_{3-j} \right){\bf w}_j,
\end{align}
with the aid of \eqref{99}. The solvability conditions are
\begin{subequations}
\begin{align}
     2\left(\lambda_3 + i\mu_3c_{g,1}(n+\mu_0)\right) + i\gamma_0\mathfrak{s}_{3,n} &= 0, \label{101a} \\
    2\gamma_0\left(\lambda_3 + i\mu_3c_{g,-1}(n+p+\mu_0)\right) + i \mathfrak{s}_{3,n+p} + i\gamma_1 \mathfrak{t}_{3,n+p} &= 0. \label{101b}
\end{align}
\end{subequations}
Using the solvability conditions \eqref{100a}-\eqref{100b} and the collision condition \eqref{27}, it can be shown \begin{align} \mathfrak{t}_{3,n+p} \equiv 0. \end{align}
As in the $p=2$ case (Section 5), the product of $\mathfrak{s}_{3,n}$ and $\mathfrak{s}_{3,n+p}$ is related to a perfect square:
\begin{align}
    \mathfrak{s}_{3,n}\mathfrak{s}_{3,n+p} = -\frac{\mathcal{S}_3^2}{\omega(n+\mu_0)\omega(n+p+\mu)},
\end{align}
where 
\begin{align}
\mathcal{S}_3 =&~ \mathcal{T}_{3,1} + \mathcal{T}_{3,2}\hat{N}_{2,2} + \mathcal{T}_{3,3}\hat{Q}_{2,2} + \mathcal{T}_{3,4}\hat{N}_{3,3}  + \mathcal{T}_{3,5}\hat{Q}_{3,3}.
\end{align}
The expressions $\mathcal{T}_{3,j}$ are functions only of $\alpha$, as are the Stokes wave corrections $\hat{N}_{2,2}$, $\hat{Q}_{2,2}$, $\hat{N}_{3,3}$, and $\hat{Q}_{3,3}$, see Appendix A. When fully expanded, $\mathcal{S}_3$ involves several hundred terms, but each term depends only on $\alpha$. The full expression of $\mathcal{S}_3$ can be found in the appropriate Mathematica notebook provided in the Data Availability Statement. The remaining calculations at this order appear in Section 6.

\section*{Data Availability Statement}

The asymptotic expressions derived in this work can be found in the following Mathematica notebooks: \emph{wwp\textunderscore isola\textunderscore  p2.nb} ($p=2$ isola in finite depth), \emph{wwp\textunderscore isola\textunderscore p2\textunderscore id.nb} ($p=2$ isola in infinite depth), and \emph{wwp\textunderscore isola\textunderscore p3.nb} ($p=3$ isola in finite depth). 

\section*{Acknowledgements} R.C. gratefully acknowledges funding from an ARCS Foundation Fellowship and from the Ruth Jung Chinn Fellowship in Applied Mathematics at the University of Washington.

\end{document}